\documentclass{aa}

\usepackage{color,graphicx}
\usepackage{txfonts}
\usepackage{natbib}
\usepackage{aalongtable}
\usepackage{amssymb}
\newcommand\Lsun{{L$_{\sun}$}}
\newcommand\Msun{{M$_{\sun}$}}
\newcommand\Rsun{{R$_{\sun}$}}
\newcommand\micron{{$\mu$m}}

\begin{document}

\title{A multi-wavelength interferometric study of the massive young
stellar object IRAS~13481-6124\thanks{Based in part on observations
with the Very Large Telescope Interferometer of the European Southern
Observatory, under program IDs 384.C-0625, 086.C-0543, 091.C-0357.}}

\author{Paul A. Boley\inst{1,2}
  \and Stefan Kraus\inst{3}
  \and Willem-Jan de~Wit\inst{4}
  \and Hendrik Linz\inst{5}
  \and Roy van~Boekel\inst{5}
  \and Thomas Henning\inst{5}
  \and Sylvestre Lacour\inst{6,7}
  \and John D. Monnier\inst{8}
  \and Bringfried Stecklum\inst{9}
  \and Peter G. Tuthill\inst{7}
}

\institute{
  Ural Federal University, Kourovka Astronomical Observatory, 51
  Lenin Ave., Ekaterinburg 620075, Russia
  \and Max Planck Institute for Radio Astronomy, Auf dem H\"ugel 69,
  Bonn 53121, Germany
  \and University of Exeter, Astrophysics Group, Stocker Road, Exeter,
  EX4~4QL, United Kingdom
  \and European Southern Observatory, Casilla 19001, Santiago 19,
  Chile
  \and Max Planck Institute for Astronomy, K\"onigstuhl 17,
  Heidelberg 69117, Germany
  \and LESIA/Observatoire de Paris, CNRS, UPMC, Universit\'e Paris
  Diderot, 5 place Jules Janssen, 92195 Meudon, France
  \and Sydney Institute of Astronomy, School of Physics, University of Sydney, NSW 2006, Australia
  \and Department of Astronomy, University of Michigan, 830 Dennison
  Building, 500 Church St., Ann Arbor, MI 48109, United States
  \and Th\"uringer Landessternwarte Tautenburg, Sternwarte 5, 07778, Tautenburg, Germany
}

\date{\today}

\abstract{We present new mid-infrared interferometric observations of
the massive young stellar object IRAS~13481-6124, using VLTI/MIDI for
spectrally-resolved, long-baseline measurements (projected baselines
up to $\sim120$~m) and GSO/T-ReCS for aperture-masking interferometry
in five narrow-band filters (projected baselines of $\sim1.8-6.4$~m)
in the wavelength range of $7.5-13$~\micron{}.  We combine these
measurements with previously-published interferometric observations in
the $K$ and $N$ bands in order to assemble the largest collection of
infrared interferometric observations for a massive YSO to date.
Using a combination of geometric and radiative-transfer models, we
confirm the detection at mid-infrared wavelengths of the disk
previously inferred from near-infrared observations.  We show that the
outflow cavity is also detected at both near- and mid-infrared
wavelengths, and in fact dominates the mid-infrared emission in terms
of total flux.  For the disk, we derive the inner radius
($\sim1.8$~mas or $\sim6.5$~AU at 3.6~kpc), temperature at the inner
rim ($\sim1760$~K), inclination ($\sim48$\degr) and position angle
($\sim107$\degr).  We determine that the mass of the disk cannot be
constrained without high-resolution observations in the
(sub-)millimeter regime or observations of the disk kinematics, and
could be anywhere from $\sim10^{-3}$ to $20$~\Msun.  Finally, we
discuss the prospects of interpreting the spectral energy
distributions of deeply-embedded massive YSOs, and warn against
attempting to infer disk properties from the SED.}

\keywords{stars: massive - techniques: interferometric - stars:
  individual: IRAS~13481-6124}

\maketitle

\section{Introduction}
\label{sec_intro}

Studies of the formation processes of massive stars have been strongly
limited by the difficulty of observing these objects at early stages.
The best evidence for an AU-scale disk around an accreting, massive
young star is delivered by the source IRAS~13481-6124.  This bright
infrared source was first identified as a young stellar object (YSO)
by \citet{Persson87}, and specifically as a massive YSO by
\citet{Chan96}, based on the flux densities measured with the IRAS
satellite.  Kinematic distance estimates \citep[which are typically
only accurate to within about a factor of 2;][]{Xu06} to this object
range from 3.1 to 3.8~kpc \citep[e.g.][]{Busfield06,Urquhart07}, and
no other distance measures are available; thus, as is often the case
with such objects, the distance is poorly known.  For the present
study, we adopt a distance of 3.6~kpc, which \citet{Fontani05}
determined kinematically from observations of rotational transitions
of CS and C$^{17}$O.  Scaling the infrared luminosity of $1.8 \times
10^5$~\Lsun{} determined by \citet{Beck91} to this distance yields a
luminosity of $\sim7.7 \times 10^4$~\Lsun{} for IRAS~13481-6124.

Observations of hydrogen recombination lines at near-infrared
wavelengths by \citet{Beck91} showed the presence of an ionizing wind
associated with IRAS~13481-6124.  From the Br$\gamma$/Pf$\gamma$ line
ratio, these authors estimated a visual extinction of $A_V = 24$~mag,
indicating that the source is fairly deeply embedded.  Interestingly,
several searches for masers \citep[e.g.][]{Scalise89,MacLeod98}, which
are often found in regions of massive star formation, failed to reveal
any maser emission associated with the source.  \citet{Beltran06}
mapped the continuum emission at 1.2~mm in the region, suggesting an
envelope mass (gas + dust) of 1470~\Msun{} (for a distance of
3.6~kpc).

Mid-infrared spectropolarimetric observations by \citet{Wright08}
revealed the presence of a sharp absorption feature at 11.2~\micron{}
in the linear polarization spectrum.  To date, this feature has only
been observed in one other source \citep[AFGL~2591;][]{Aitken88}, and
both \citet{Aitken88} and \citet{Wright08} interpret the feature as
possibly arising from the presence of crystalline olivine, which
could form as a result of dust processing in the circumstellar
environment.  The apparent rarity of this feature, as noted by
\citet{Wright08}, could indicate that IRAS~13481-6124 is currently in
a very short-lived phase of the evolutionary sequence of massive star
formation.  We note, however, that alternative explanations could be
that this feature only occurs if the source has a particular geometry,
or if viewed under particular inclination angles.

Interferometric measurements of the source at near-infrared
wavelengths were performed by \citet{Kraus10}.  The $K$-band
($\lambda=2.2$~\micron) image reconstructed from these interferometric
observations shows an elongated structure, which is oriented
perpendicular to the large-scale outflow seen in Spitzer Space
Telescope images from the GLIMPSE survey \citep{Churchwell09}, and
more clearly in the continuum-subtracted H$_2$ images
($\lambda=2.122$~\micron{}) presented by \citet{Caratti15}.
Spatially-resolved $Q$-band ($\lambda=20$~\micron) imaging by
\citet{Wheelwright12} shows an elongated structure roughly along the
outflow direction.  Radiative transfer modeling of the spectral energy
distribution (SED) and near-infrared interferometric data by
\citet{Kraus10} suggest that IRAS~13481-6124 hosts a circumstellar
disk, and suggest a total luminosity of $4\times10^4$~\Lsun{}
\citep[i.e., about half the value derived from the IRAS measurements
by][]{Beck91}.

Most recently, $K$-band integral-field spectroscopic measurements by
\citet{Stecklum12} showed a photocenter shift of $\sim1$~AU in the
Br$\gamma$ line along the same position angle as the bow shock
reported by \citet{Kraus10}, which confirms the association of the
parsec-scale outflow with the MYSO.  Furthermore, these integral-field
measurements also show evidence for (possibly non-Keplerian) rotation
perpendicular to the outflow direction, providing yet another
indication of a circumstellar disk in this system.

In this work, we consider the $K$-band and $N$-band
($\lambda=8-13$~\micron) interferometric measurements presented by
\citet{Kraus10} and \citet{Boley13}, respectively, together with new
$N$-band interferometric measurements of the source, obtained using
both long-baseline and aperture-masking interferometry. We use
geometric models of the near-/mid-infrared interferometric
observations, and probe the source on spatial scales (given by
$\sim\lambda / 2B$, where $\lambda$ is the wavelength, and $B$ is the
projected baseline of the interferometric observations) of about
10--1000~AU.  We focus on the wavelength behavior of the fits, which
we examine in the context of the temperature profile of the
circumstellar disk.

\section{Observations}

\subsection{Long-baseline interferometry with VLTI}

\begin{table*}\footnotesize
\begin{center}
\caption{Log of new MIDI observations of IRAS 13481-6124}
\label{tab_obslog}
\begin{tabular}{l l c c c c c}
\hline \hline
Date/time (UTC) & Telescopes & Proj. baseline & Position angle & Avg. visibility & Dispersive & Program ID \\
& & (m) & (deg.) & & element & \\
\hline
 2010-02-14~~08:39 & E0-G0 & $15.8$ & $71.0$ & $0.42$ & PRISM & 384.C-0625(A)\\
 2011-03-11~~03:55 & G1-I1 & $44.8$ & $9.4$ & $0.20$ & PRISM & 086.C-0543(B)\\
 2011-03-11~~04:06 & G1-I1 & $44.7$ & $11.5$ & $0.20$ & PRISM & 086.C-0543(B)\\
 2011-03-11~~05:28 & G1-H0 & $57.8$ & $174.7$ & $0.15$ & PRISM & 086.C-0543(B)\\
 2011-03-11~~07:49 & D0-I1 & $82.3$ & $104.8$ & $0.09$ & PRISM & 086.C-0543(B)\\
 2011-03-11~~08:28 & H0-I1 & $36.3$ & $155.2$ & $0.21$ & PRISM & 086.C-0543(A)\\
 2011-03-13~~05:51 & G1-K0 & $79.3$ & $19.0$ & $0.13$ & PRISM & 086.C-0543(C)\\
 2011-03-31~~04:52 & D0-I1 & $79.9$ & $83.2$ & $0.09$ & PRISM & 086.C-0543(B)\\
 2011-03-31~~06:16 & D0-I1 & $82.1$ & $101.5$ & $0.08$ & PRISM & 086.C-0543(B)\\
 2011-05-03~~05:44 & G1-I1 & $36.4$ & $66.9$ & $0.22$ & PRISM & 086.C-0543(A)\\
 2012-02-24~~06:42 & D0-I1 & $78.5$ & $76.6$ & $0.12$ & PRISM & 086.C-0543(C)\\
 2012-03-29~~02:53 & A1-I1 & $102.4$ & $41.3$ & $0.14$ & PRISM & 086.C-0543(A)\\
 2012-03-29~~03:03 & A1-I1 & $102.8$ & $43.6$ & $0.12$ & PRISM & 086.C-0543(A)\\
 2012-04-01~~07:00 & A1-I1 & $104.5$ & $100.1$ & $0.09$ & PRISM & 086.C-0543(A)\\
 2012-04-01~~09:02 & G1-K0 & $59.7$ & $60.2$ & $0.18$ & PRISM & 086.C-0543(C)\\
 2012-04-01~~09:11 & G1-K0 & $58.6$ & $61.6$ & $0.18$ & PRISM & 086.C-0543(C)\\
 2013-04-30~~04:13 & UT1-UT4 & $123.3$ & $68.3$ & $0.06$ & GRISM & 091.C-0357(B)\\
 2013-04-30~~04:27 & UT1-UT4 & $122.1$ & $71.3$ & $0.06$ & GRISM & 091.C-0357(B)\\
 2013-05-01~~06:21 & UT2-UT3 & $34.5$ & $72.8$ & $0.20$ & GRISM & 091.C-0357(A)\\
\hline

\end{tabular}
%\tablefoot{Besides the (Chapter~\ref{chap_survey}).}
\end{center}
\end{table*}

New observations of IRAS~13481-6124 were conducted in the $N$ band
(8--13~\micron{}) using the mid-infrared interferometric instrument
MIDI \citep{Leinert03} at the Very Large Telescope Interferometer
(VLTI) of the European Southern Observatory (ESO) in 2010--2013.  The
MIDI instrument is a two-telescope beam combiner which measures
spectrally-dispersed correlated and total flux, with either a grism
($\lambda / \Delta \lambda \approx 230$) or a prism ($\lambda / \Delta
\lambda \approx 30$) serving as the dispersive element.  Using MIDI,
light can be combined simultaneously from either two of the 8-m Unit
Telescopes (UTs) or two of the 1.8-m Auxiliary Telescopes (ATs).

The observations presented in this paper were taken in the HIGHSENS
mode, meaning the correlated and total flux measurements are made
separately.  We summarize the interferometric measurements in
Table~\ref{tab_obslog}, where we show the time the fringe track was
started, the telescopes/interferometric stations used, the projected
baseline and position angle, as well as the average visibility,
dispersive element and ESO program ID.  Each interferometric
measurement was preceded and/or followed by a calibration measurement
of a bright star of known brightness and diameter, taken from the
van~Boekel mid-infrared calibrator star database \citep{vanBoekelPhD}.
The observations in 2013 were obtained using the UTs with the grism as
the dispersive element, while the observations in 2010--2012 used the
ATs and the lower-resolution prism.

Data reduction was performed using version 2.0Beta1 (8 Nov. 2011) of
the MIA+EWS package \citep{Jaffe04}.  We used the fringe pattern of
the individual calibrator stars to create a median mask for each
night/baseline configuration, which was then used for extracting the
correlated flux from the science measurements.  Calibrated correlated
fluxes (in Jy) were computed using the measurements of the calibrator
stars, and the uncertainty in the correlated flux was taken as the
root-mean-square sum of the random error derived from MIA+EWS and the
systematic error derived from the calibrator measurements.

\begin{figure*}
  \begin{center}
    \includegraphics[width=140mm]{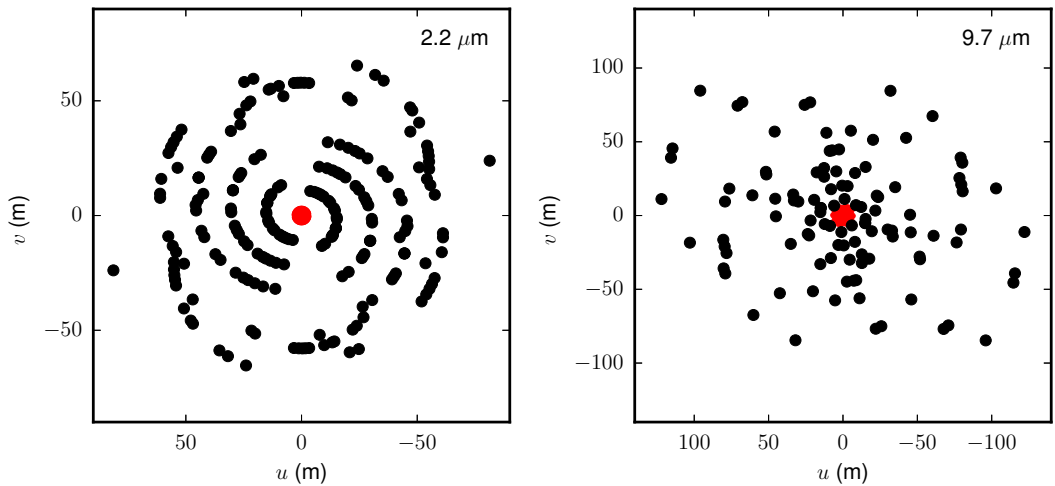}
    \caption{$uv$ coverage of the interferometric measurements.  The
      left panel shows the coverage of the near-infrared measurements,
      while the right panel shows the coverage of the mid-infrared
      measurements.  Black points show the long-baseline VLTI
      observations (AMBER and MIDI), while the speckle-interferometry
      and aperture-masking observations (NTT and GSO/T-ReCS) are shown
      in red.}
    \label{fig_uvplot}
  \end{center}
\end{figure*}

\begin{figure}
  \begin{center}
    \includegraphics[width=85mm]{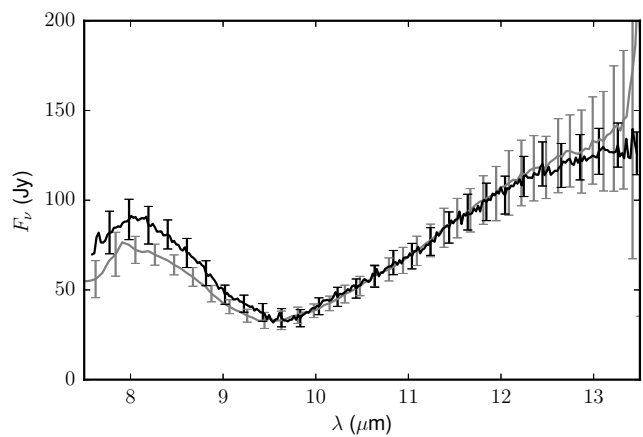}
    \caption{Total $N$-band spectrum of IRAS~13481-6124.  The gray
      line shows the spectrum observed with the prism \citep{Boley13};
      the black line shows the spectrum measured with the grism.}
    \label{fig_spectra}
  \end{center}
\end{figure}

Finally, we also make use of the 36 MIDI observations presented as
part of the VLTI/MIDI MYSO Survey \citep{Boley13}.  The observations
presented in that work were reduced in an identical fashion as in the
present work.  The projected baselines of this combined MIDI data set
range from $\sim9$ to $\sim128$~m; we show the resulting $uv$ coverage
in the right-hand panel of Fig.~\ref{fig_uvplot}.  As in the work by
\citet{Boley13}, we derive spectrally-dispersed visibilities from the
correlated fluxes by dividing the correlated flux spectra by the
median spectrum measured with the UTs.  For the observations with the
prism, we use the same total spectrum (also obtained with the prism)
as \citet{Boley13}; for the observations with the grism, we use the
median of the total flux spectra obtained with the grism on 2013-04-30
and 2013-05-01 (see Fig.~\ref{fig_spectra}).  OIFITS files of the MIDI
data are available upon request.

\subsection{Aperture-masking interferometry with Gemini-South}

Short-baseline interferometric observations of IRAS~13481-6124 were
also performed using the aperture-masking technique implemented at the
Thermal-Region Camera Spectrograph \citep[T-ReCS;][]{deBuizer05} at
the Gemini South Observatory (GSO) on May~7, 2007 (program ID
GS-2007A-Q-38).  This technique works by apodizing the pupil with a
mask with several holes in it, thus transforming the single large
telescope mirror into an array of smaller sub-apertures.  Images
formed are now Fizeau interference patterns, and the complex
visibilities corresponding to each baseline passed by the mask may now
be extracted using Fourier transform techniques. Further details of
this technique are described by \citet{Tuthill00}, while an
interesting adaptation of the technique to mid-infrared
interferometric data is given by \citet{Monnier04}.  A recent example
of the application of this technique can be found in the work of
\citet{Vehoff10}, who used T-ReCS observations similar to those
presented here to reconstruct a diffraction-limited image at a
wavelength of 11.7~\micron{} of the MYSO NGC~3603~IRS~9A.

A set of rapid exposures with a total on-source integration time of
131~s were performed in the five narrow-band filters Si1, Si2, Si3,
Si5 and Si6, which have central wavelengths of 7.7, 8.7, 9.7, 11.7 and
12.4~\micron{}, respectively, and widths in the range of
$\sim0.7$--$1.2$~\micron{} (bandpasses overplotted in
Fig.~\ref{fig_miditrecs}).  The raw visibility quantities extracted
($V^2$ and closure phase) are calibrated for the effects of the
telescope-atmosphere transfer function with the use of contemporaneous
observations of a nearby point-source reference star.  The
aperture-masking observations presented here measure the visibility
amplitude with projected baselines of 1.8--6.4~m; the Fourier coverage
in the $uv$ plane is concentrated in the red core of the right panel
of Fig.~\ref{fig_uvplot}.  The closure phases (not presented in this
work) were generally small ($\la10$\degr), as expected from a
predominantly point-symmetric object.  OIFITS files of the T-ReCS
observations are available upon request.

\subsection{Previously-published $K$-band observations}

IRAS~13481-6124 was observed in the $K$ band with the three-telescope
interferometric instrument AMBER at the VLTI in 2008--2009 (program
IDs 081.C-0272, 083.C-0236, PI G.~Weigelt; 083.C-0621, PI S.~Kraus),
and speckle-interferometry observations were made with the 3.6~m ESO
New Technology Telescope (NTT) on February 3, 2009 (program ID
082.C-0223, PI S.~Kraus).  These data were presented by
\citet{Kraus10}, and we refer to that publication for further details
about the observations and data reduction, however we reiterate the
most relevant characteristics of the data here.

The projected baselines of the AMBER measurements range from 11 to
85~m, while the NTT bispectrum speckle measurements provide projected
baselines up to $\sim3$~m.  The AMBER data consist of
spectrally-resolved visibility amplitudes and closure phases across
the $K$ band (1.95--2.55~\micron{}) with a spectral resolution
$\lambda/\Delta \lambda \approx 35$, while the NTT speckle data
consist of just visibility amplitudes at the center of the $K$ band
(2.2~\micron).  We show the $uv$ coverage in the left panel of
Fig.~\ref{fig_uvplot}, where the VLTI/AMBER measurements are shown in
black, and the NTT speckle data are shown in red.

\section{Results}
\label{sec_results}

\begin{figure*}[t]
  \begin{center}
    \includegraphics[width=170mm]{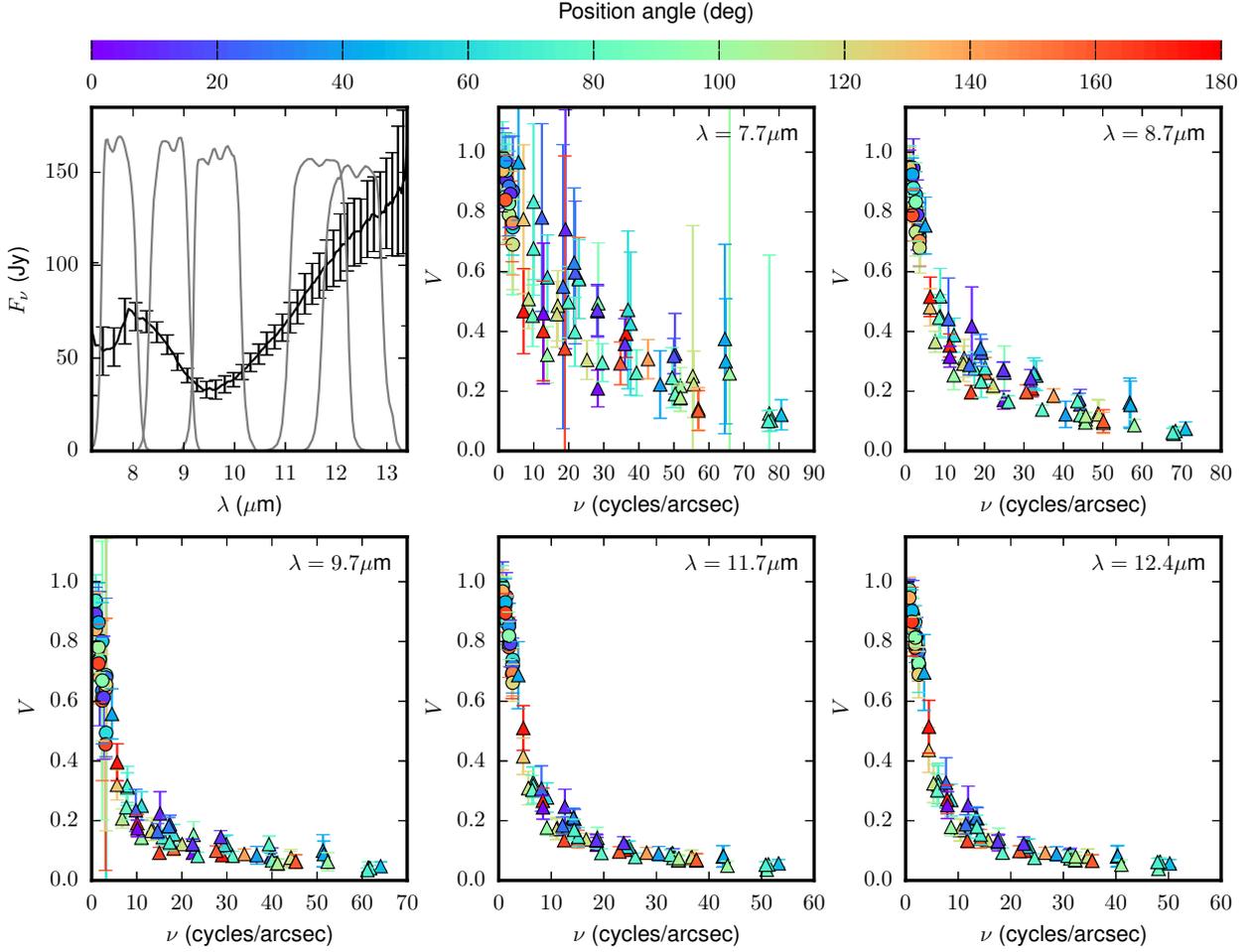}
    \caption{\emph{Top left panel:} Total $N$-band spectrum observed
      with MIDI (black) and the transmission curves for the five
      filters used for the T-ReCS aperture-masking
      measurements. \emph{Remaining panels:} The visibility amplitude
      $V$ as a function of spatial frequency $\nu$ in each of the five
      filters.  The T-ReCS measurements are shown as circles, while
      the MIDI measurements, averaged over each corresponding filter,
      are shown as triangles.  The color of each data point shows the
      position angle (measured east of north).}
    \label{fig_miditrecs}
  \end{center}
\end{figure*}

The results of the MIDI and T-ReCS observations are shown in
Fig.~\ref{fig_miditrecs}.  In the top left panel, as a black line, we
show the total $N$-band spectrum of the source together with the
transmission curves of the five filters used for the T-ReCS
observations, overlaid in gray.  The spectrally-resolved MIDI
visibilities were averaged over these filter curves to create values
which are directly comparable to the T-ReCS measurements.  The
remaining five panels of Fig.~\ref{fig_miditrecs} show the visibility
amplitude $V$ as a function of spatial frequency $\nu = B / \lambda$
in each of the five filters, where the T-ReCS data are shown as
circles, and the MIDI data are shown as triangles; the color of each
data point shows the position angle (measured east of north) of the
measurement.  \textbf{The increased level of noise seen in the
observations at $\lambda=7.7$~\micron{} is due to the much lower
atmospheric transmission at this wavelength, which is near the edge of
the $N$ band.}

First, we note that the transition from the visibilities measured with
MIDI to those measured with T-ReCS is smooth and continuous.  This
confirms that the absolute calibration between these two very
different techniques and instruments is consistent, and that the
long-baseline observations are directly comparable to those obtained
with the aperture-masking technique.

Second, the dependence of the visibility on baseline and wavelength
shows that the infrared emission can be approximately described by two
components: an extended component causes the visibility to drop
quickly at short baselines (spatial frequencies $\la$10
cycles/arcsec), and a compact component then makes the visibility
curve flatten off and decrease more slowly towards longer baselines.
The spatial frequency at which the visibility curve flattens is a
measure of the size of the extended component, and the visibility
amplitude at this point equals the fraction of the total flux that is
contributed by the compact component.

Finally, a distinct emission feature was noted at $\sim11.2$~\micron{}
in the $N$-band spectrum of polarized light taken with TIMMI2
\citep{Wright08}. We note in passing that our MIDI data show no
obvious feature at this wavelength, neither in the visibilities nor in
the correlated fluxes. This specifically includes the few GRISM data
points, which have a higher spectral resolution.

\section{Analysis}
\label{sec_analysis}

In this section, in order to perform a reliable analysis of the infrared
interferometric observations, we use a series of progressively-complicated
approaches to analyze the data.  First, we evaluate our assumptions about the
near- and mid-infrared dust extinction in Sec.~\ref{sec_ext}.  We follow this
with geometric fits to the visibilities in Sec.~\ref{sec_visfits}.  Finally, in
Sec.~\ref{sec_cflux_models}, we elaborate on this approach to include the flux
information contained in the mid-infrared observations, both for geometric
models (Sec.~\ref{sec_tgm_cflux}) and for radiative transfer models
(Sec.~\ref{sec_rt_models}).

\subsection{Near- and mid-infrared extinction}
\label{sec_ext}

IRAS~13481-6124, like most MYSOs, is heavily affected by extinction.
Using the ratio of the Br$\gamma$ and Pf$\gamma$ Hydrogen emission
lines, \citet{Beck91} estimated a total (visual) extinction of
$A_V\approx24$~mag for this source.  The deep silicate absorption
feature seen in the total $N$-band spectrum (Fig.~\ref{fig_spectra})
further attests to the high amount of extinction towards the object,
even at mid-infrared wavelengths.

Since we are considering observations at near- and mid-infrared
wavelengths in this work, it is necessary to consider the extinction
over a wavelength range of $\sim1-13$~\micron{}~-- a range for which
it has traditionally been very difficult to characterize extinction
laws (see, for example, the discussions by \citet{Lutz99} and
\citet{Fritz11}).  We find that the shape of the silicate feature in
this object is poorly reproduced by the synthetic extinction curves of
\citet{WD01} and the empirical extinction curve of \citet{Fritz11}.
However, using opacities for amorphous silicates measured in the
laboratory by \citet{Dorschner95}, \citet{Boley13} showed that the
absorption feature in the $N$-band correlated flux spectra of
IRAS~13481-6124 is well-reproduced by a mixture consisting primarily
of ``small'' grains ($r=0.1$~\micron) of pyroxene glass (in the form
of MgFeSiO$_3$) and olivine glass (in the form of MgFeSiO$_4$), with
relative mass abundances of 63\% and 36\%, respectively, and less than
1\% of the mass in ``large'' ($r=1.5$~\micron) grains.  Furthermore,
it was also shown in this work that the shape of the absorption
feature in the correlated flux spectra is independent of the spatial
scale probed, on projected baselines from 9 to 128~m.

These results from \citet{Boley13} characterize the contribution of
silicates to the line-of-sight extinction in this particular object,
however they cannot provide an estimate of the contribution from
carbonaceous grain species, which can contribute significantly (or
even dominantly, depending on wavelength) to the continuum opacities
in interstellar extinction laws.  Furthermore, due to the lack of any
strong features from most types of carbonaceous species at infrared
wavelengths, it is essentially impossible to directly constrain this
important dust component for deeply-embedded objects using
conventional methods.

\begin{figure}
  \begin{center}
    \includegraphics[width=85mm]{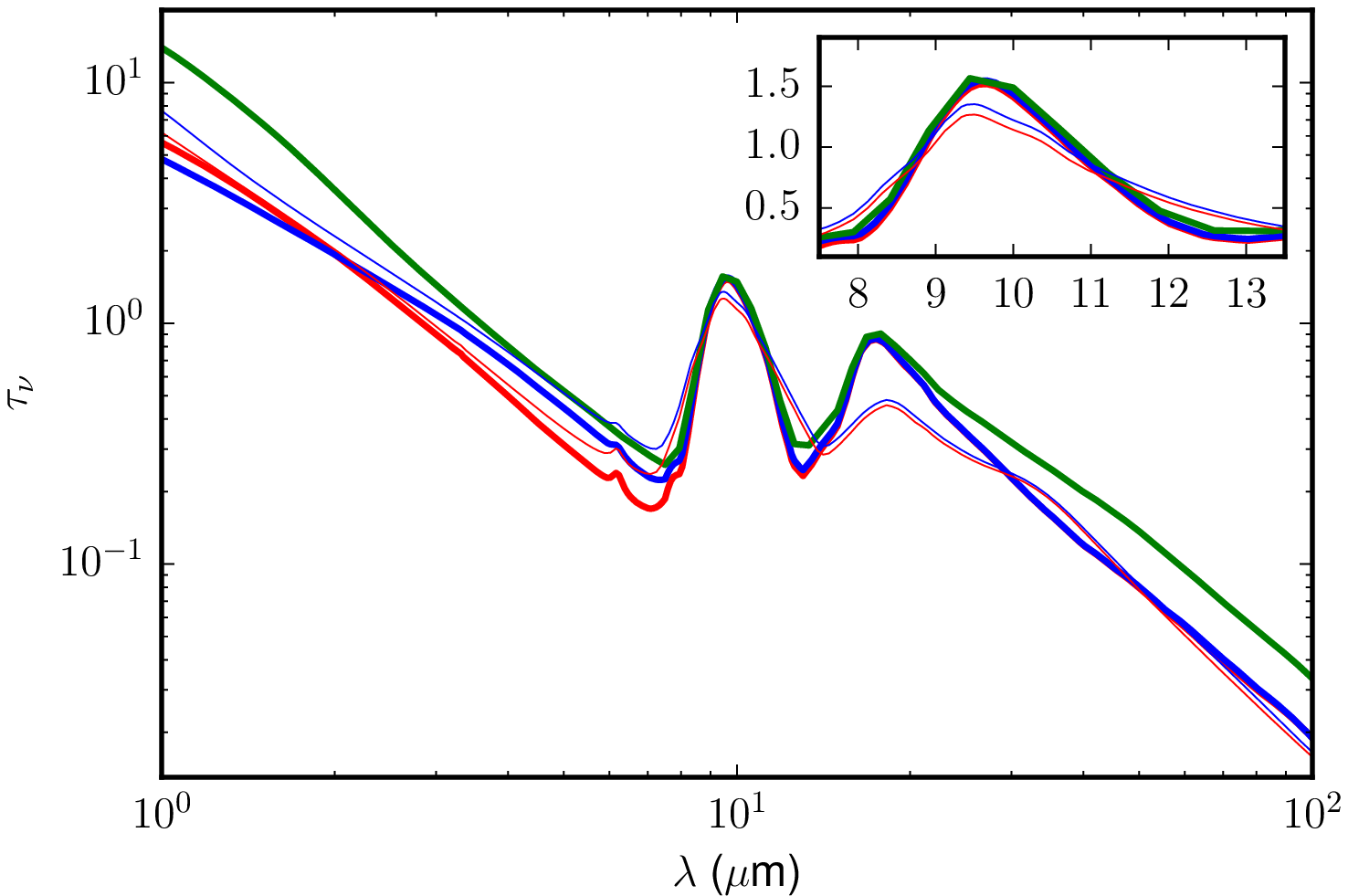}
    \caption{Derived optical depth of the foreground extinction,
      for different types of carbonaceous grains and size
      distributions.  The thick red and blue curves utilize the
      Milky-Way size distributions from \citet{WD01} for the $R_V=3.1$
      and $R_V=5.5$ extinction laws, respectively, using PAH and
      graphite grains.  The thick green curve utilizes the MRN size
      distribution with graphite grains.  All three curves include the
      silicate contribution derived by \citet{Boley13}.  For
      comparison, the synthetic extinction curves of \citet{WD01},
      which differ in their silicate content, are shown as thin red
      ($R_V=3.1$) and blue ($R_V=5.5$) curves.}
    \label{fig_tau}
  \end{center}
\end{figure}

As we cannot measure this contribution from carbonaceous grains
directly, a reasonable approach to estimate the total foreground
extinction towards IRAS~13481-6124 is to simply \emph{adapt} the
contribution of carbonaceous grains from a synthetic extinction curve,
and combine this with the determination of the silicate component by
\citet{Boley13}.  In Fig.~\ref{fig_tau}, we show the extinction
optical depth $\tau_\nu(\lambda)$ thus derived, where we have taken
the same carbonaceous mixture used by \citet{WD01} (i.e., graphite
grains from \citet{Laor93} and PAH grains from \citet{Li01}).  We used
the Milky-Way size distribution from \citet{WD01}, with two sets of
parameters from Table~1 of that work, describing the $R_V=3.1$
($b_C=60$~ppm) and $R_V=5.5$ ($b_C=30$~ppm, case ``A'') extinction
laws.  For comparison, we also show the optical depth derived using
instead the graphite grains of \citet{Li01} with the often-used
``MRN'' \citep[$dn \propto a^{-3.5}da$, $a_\mathrm{min}=5$~nm,
  $a_\mathrm{max}=250$~nm;][]{MRN77} size distribution and a mass
ratio of silicates to graphite of 60:40, as well as the $R_V=3.1$ and
$R_V=5.5$ Milky-Way synthetic extinction curves from \citet{WD01}.

We note that while the three different extinction curves we derive
show essentially identical values for $\tau_\nu$ in the $N$-band
(inset of Fig.~\ref{fig_tau}), the scatter between all three curves at
near-infrared wavelengths is large.  Furthermore, at near-infrared
wavelengths $\tau_\nu$ is significantly larger than unity.  In terms
of magnitudes of extinction ($A_\lambda=2.5 \log_{10}(e)
\tau_\nu(\lambda)$),  $A_K$ ranges from 1.8 to 3.1~mag, and $A_J$
ranges from 4.0 to 10~mag.  This is of particular importance, as it
implies that the derived extinction at near-infrared wavelengths is
highly uncertain, and that reliable dereddened flux levels (which
scale as $e^{\tau_\nu(\lambda)}$) can not be obtained at these
wavelengths.  This matter is addressed in more detail in
Section~\ref{sec_dis_RT}.

\subsection{Model fits to the visibilities}
\label{sec_visfits}

As noted in Section~\ref{sec_results}, the observed visibilities of
IRAS~13481-6124 show evidence for two distinct components.
\citet{Kraus10} examined the $K$-band AMBER/NTT visibilities in terms
of several types of two-component geometric models, all of which
included an extended, one-dimensional Gaussian ``halo'' as one
component; the second (two-dimensional) component was either a uniform
disk, a thin ring, a compact Gaussian, or a temperature-gradient disk
model\footnote{See Appendix~\ref{sec_K_refit} for a correction to the
geometric fits presented by \citet{Kraus10}.}.  In the following
subsections, we further explore the geometry of the system in the
context of the additional information provided by the mid-infrared
observations.

\subsubsection{Gaussian models}
\label{sec_analysis_gauss}

As the Gaussian model provides good fits to the near-infrared data,
has fewer parameters than the parameterized disk model, and is
conceptually very simple, we begin with this type of model to examine
the entirety of the interferometric data.  Specifically, we fit the
``2D1D'' model from \citet{Boley13} to each wavelength separately;
this model consists of a two-dimensional Gaussian (characterized by a
full-width-at-half-maximum size FWHM$_\mathrm{2D}$, position angle
$\phi$ and inclination angle $i$) and a one-dimensional Gaussian
(size FWHM$_\mathrm{1D}$), both centered at the origin.  The flux
ratio between these two components is given by
$F_\mathrm{2D}/F_\mathrm{1D}$.  The best-fit parameters were derived
with a grid search followed by a downhill-simplex minimization of
$\chi^2$; uncertainties in the best-fit values were derived using the
same Monte-Carlo approach described by \citet{Boley13}.

\begin{figure*}
  \begin{center}
    \includegraphics[width=170mm]{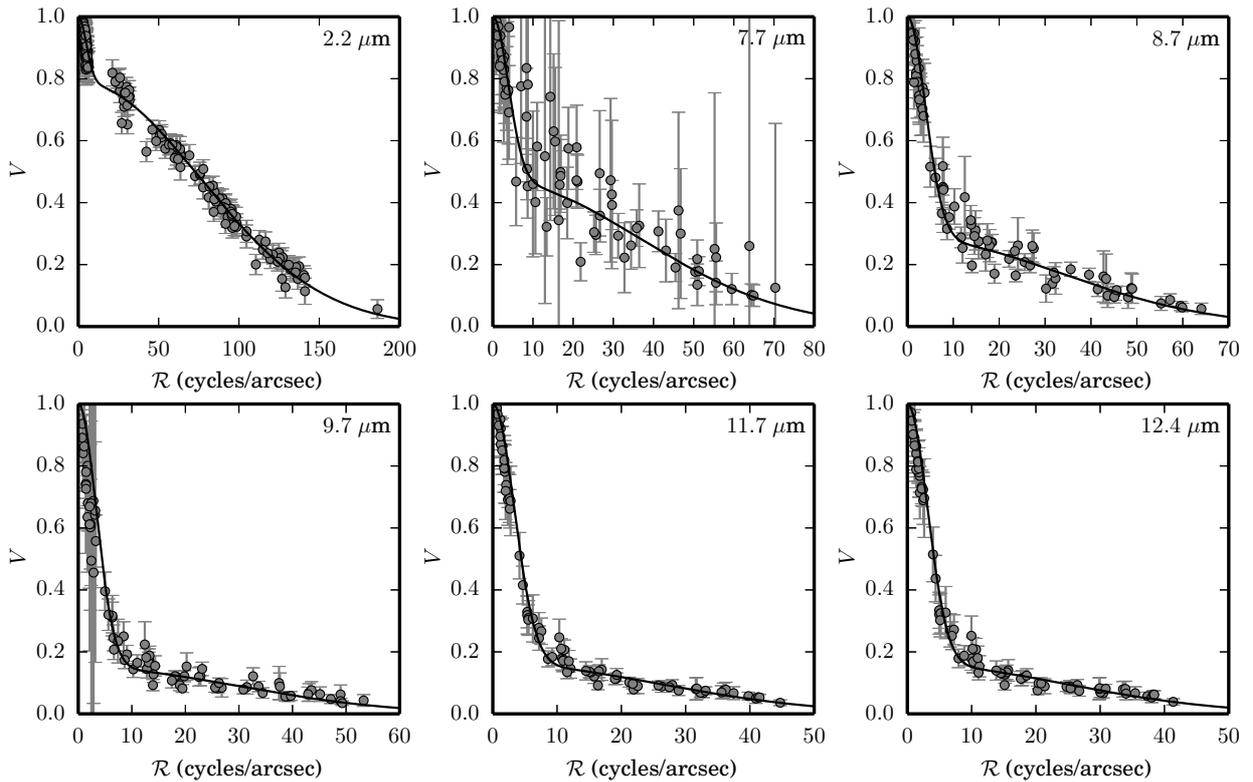}
    \caption{Gaussian model fits to the visibilities at each
      wavelength, plotted as a function of deprojected spatial
      frequency.  See Table~\ref{tab_gaussfit} for the fit parameters.
      Circles and error bars mark the observations;
      while the model fits to each wavelength are shown as a solid
      curve.}
    \label{fig_gaussfit}
  \end{center}
\end{figure*}

\begin{table*}
\begin{center}
  \caption{Fit parameters for 2D1D model}
  \label{tab_gaussfit}
  \begin{tabular}{c c c c c c}
    \hline \hline
    Wavelength & $F_\mathrm{2D}/F_\mathrm{1D}$ & FWHM$_\mathrm{2D}$ & FWHM$_\mathrm{1D}$ & $\phi$ & $i$ \\
    \multicolumn{1}{c}{(\micron)} & & \multicolumn{1}{c}{(mas)} & \multicolumn{1}{c}{(mas)} & \multicolumn{1}{c}{(deg)} & \multicolumn{1}{c}{(deg)} \\
    \hline
    2.2 & $3.75\pm 0.16$ & $4.93\pm 0.062$ & $76.8\pm 3.0$ & $112.\pm 1.7$ & $45.0\pm 1.2$ \\
    7.7 & $0.894\pm 0.13$ & $10.3\pm 0.87$ & $99.3\pm 15.$ & $120.\pm 13.$ & $45.3\pm 8.8$ \\
    8.7 & $0.398\pm 0.024$ & $11.3\pm 0.56$ & $96.5\pm 6.6$ & $114.\pm 8.1$ & $43.7\pm 5.5$ \\
    9.7 & $0.176\pm 0.012$ & $12.8\pm 1.1$ & $113.\pm 7.8$ & $131.\pm 13.$ & $43.1\pm 7.4$ \\
    11.7 & $0.191\pm 0.013$ & $14.6\pm 1.1$ & $114.\pm 5.9$ & $130.\pm 11.$ & $42.2\pm 6.6$ \\
    12.4 & $0.190\pm 0.014$ & $15.3\pm 1.2$ & $121.\pm 6.4$ & $134.\pm 14.$ & $42.1\pm 7.0$ \\
    \hline
  \end{tabular}
\end{center}
\end{table*}

We tabulate the results of this fitting procedure for each wavelength
in Table~\ref{tab_gaussfit}, and show the observed and model
visibilities in Fig.~\ref{fig_gaussfit}.  The visibilities are plotted
in terms of the deprojected spatial frequency $\mathcal{R}$,
\begin{equation}
\label{eq_deproj}
\mathcal{R} = \sqrt{u_\phi+v_\phi^2\cos^2{i}},
\end{equation} where
\begin{eqnarray}
u_\phi & = & u\cos\phi-v\sin\phi \\
v_\phi & = & u\sin\phi+v\cos\phi
\end{eqnarray}
and $u$ and $v$ are the spatial coordinates of the interferometric
observations.  The inclination angle $i$ and position angle $\phi$ are
taken from the best-fit values for each wavelength in
Table~\ref{tab_gaussfit}.

We find that the two-dimensional parameters derived from the
2.2~\micron{} data are essentially identical to those derived by
\citet{Kraus10} for the same data set.  However, in contrast to the
present work, the relative flux and size of the one-dimensional
Gaussian component in their model were not free parameters: we find a
FWHM size of 77~mas (compared to their fixed value of 108~mas), and a
relative flux of the one-dimensional Gaussian of 27\% (compared to
their fixed value of 15\%).

The orientation derived from the mid-infrared data matches that
derived from the near-infrared observations to within
$\sim10$--$20$\degr, which \citet{Kraus10} showed to be perpendicular
to the large-scale outflow, and interpreted as originating from a
circumstellar disk.  Thus, for this particular MYSO, the mid-infrared
emission of the compact component is clearly aligned with the
expected orientation of the disk\footnote{N.B., this is not always the
case.  See the discussion of individual sources in the VLTI/MIDI MYSO
Survey \citep{Boley13}.}.  Furthermore, the widths of the
two-dimensional Gaussians in the fit increase with wavelength.
Similar behavior was also seen by \citet{Kraus10} in the
spectrally-resolved near-infrared data, and is indicative of a
temperature gradient: i.e., the source looks larger at longer
wavelengths (lower temperatures).  We thus confirm that the
temperature gradient detected by \citet{Kraus10} is also apparent at
$N$-band wavelengths.

\subsubsection{Temperature-gradient disk models}
\label{sec_tgm_vis}

The Gaussian fits presented in Section~\ref{sec_analysis_gauss}
provide a simple way to examine the general source morphology at
different wavelengths.  However, such Gaussian fits are an ad-hoc
solution to the problem of fitting visibilities, rather than a
physically-motivated model with meaningful physical parameters.

An attractive alternative to the Gaussian models is a model consisting
of a geometrically-thin, optically-thick (i.e., emitting as a black
body) disk with a temperature distribution parameterized by a power
law, i.e.
\begin{equation}
  \label{eq_Tlaw}
  \begin{array}{lr}
    T(r) = T_\mathrm{in} \left(\frac{r_\mathrm{in}}{r} \right)^p & r_\mathrm{in} < r \leq r_\mathrm{out},
  \end{array}
\end{equation}
where $T_\mathrm{in}$ is the temperature at the inner radius
$r_\mathrm{in}$ of the disk.  Such a temperature gradient follows from
several theoretical studies of disk structure \citep[for an overview,
  see the textbook by][and references therein]{Hartmann09}, and has
been applied to T~Tauri stars \citep{Eisner05,Vural12}, Herbig stars
\citep{Eisner07,Benisty11,Ragland12,Chen12,Kraus13,Vural14a,Vural14b},
MYSOs \citep{Kraus10,Boley12} and other objects
\citep[e.g.][]{Malbet05,Kreplin12,Kraus12,Wang12}.

This model provides a single physical structure which can emit at a
variety of wavelengths and scales.  For additional information on
constructing the visibility function for such a disk, we refer to the
work by \citet{Eisner07}.  As in Section~\ref{sec_analysis_gauss}, we
also include an extended component in the form of a one-dimensional
Gaussian ``halo.''  However, as this extended component is not well
probed by our data, we do not attempt to ascribe a predefined
wavelength dependence to the halo parameters.  Thus, the relative flux
and size of the halo are fit independently for all wavelengths, while
the disk parameters are fit simultaneously for all wavelengths.  

\begin{figure*}
  \begin{center}
    \includegraphics[width=170mm]{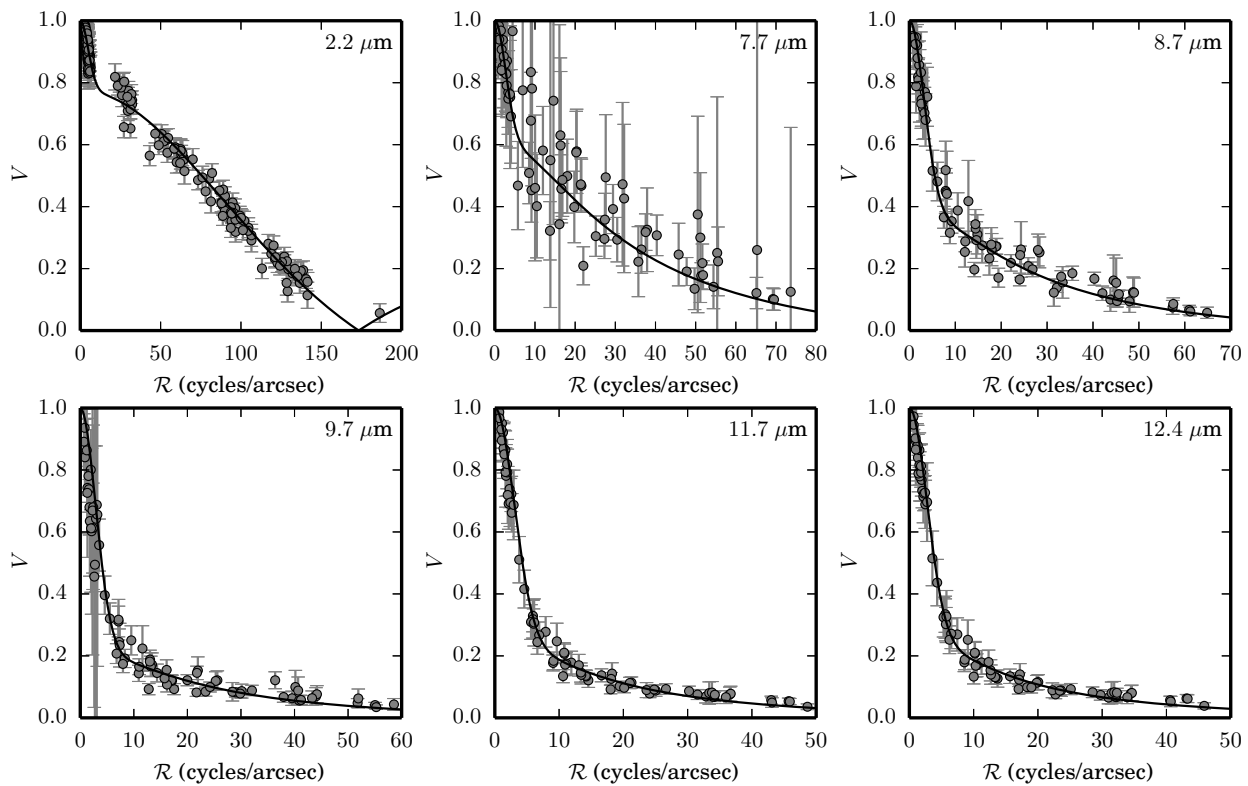}
    \caption{Temperature-gradient disk model fits to the visibilities
      for a disk with $T_\mathrm{in}=1500$~K (fixed; Model~1), plotted
      as a function of deprojected spatial frequency.  See
      Table~\ref{tab_tgm} (Model 1) for the fit parameters.  Circles and error
      bars mark the observations, while the model fits to each
      wavelength are shown as a solid curve.}
    \label{fig_tgm_vis}
  \end{center}
\end{figure*}

The results of fitting this model for a \emph{fixed} inner temperature
of the disk of $T_\mathrm{in}=1500$~K (approximating the evaporation
temperature of silicate grains) to the $K$- and $N$-band visibility
amplitudes are shown in Fig.~\ref{fig_tgm_vis} and in
Table~\ref{tab_tgm} (Model~1).

The quality of the fit is good ($\chi^2_\mathrm{r}=0.613$).  The disk
position angle and inclination angle derived are similar to those
found for the Gaussian models, and perpendicular to the large-scale
outflow shown by \citet{Kraus10} and \citet{Caratti15}.  However, the
inner temperature of the disk cannot be determined by fitting the
visibilities alone.  For example, if we allow $T_\mathrm{in}$ to be a
free parameter when fitting only the visibilities, we do find a
formally ``better'' fit (not shown; $\chi^2_\mathrm{r}=0.538$); but
with an inner disk temperature of $T_\mathrm{in}=7490$~K at a radius
of $R_\mathrm{in}=1.02$~mas (3.6~AU at a distance of 3.6~kpc), and a
total integrated power output of $\sim5\times10^6$~\Lsun{}, such a
model is clearly unphysical, underlying our conclusion that fitting
\emph{only} the visibilities or \emph{only} the SED with
(semi-)physical models can lead to nonsensical results (for
comparison, the total luminosity for this source derived by the RMS
survey \citep{Urquhart07} is lower by a factor of $\sim10^2$).

\begin{table}
\begin{center}
  \caption{Fit parameters for temperature-gradient disk models}
  \label{tab_tgm}
  \begin{tabular}{l c c}
    \hline \hline
    Parameter & Model~1 & Model~2 \\
    \hline
    $T_\mathrm{in}$ (K) & 1500$^\dagger$ & 1760 \\
    $r_\mathrm{in}$ (mas) & 1.55 & 1.78 \\
    $p$ & 0.616 & 0.840 \\
    $\phi_\mathrm{disk}$ (deg) & 112 & 107 \\
    $i_\mathrm{disk}$ (deg) & 42.3 & 47.6 \\
    $F_\mathrm{disk}/F_\mathrm{halo}$~~ & & \\
    ~~~~~~$\lambda=2.2$~\micron & 3.45 & 3.10 \\
    ~~~~~~$\lambda=7.7$~\micron & 1.61 & 0.750 \\
    ~~~~~~$\lambda=8.7$~\micron & 0.644 & 0.370 \\
    ~~~~~~$\lambda=9.7$~\micron & 0.279 & 0.250 \\
    ~~~~~~$\lambda=11.7$~\micron & 0.347 & 0.246 \\
    ~~~~~~$\lambda=12.4$~\micron & 0.369 & 0.246 \\
    FWHM$_\mathrm{halo}$ (mas) & & \\
    ~~~~~~$\lambda=2.2$~\micron & 76.8 & 69.9 \\
    ~~~~~~$\lambda=7.7$~\micron & 142 & 61.2 \\
    ~~~~~~$\lambda=8.7$~\micron & 120 & 84.1 \\
    ~~~~~~$\lambda=9.7$~\micron & 132 & 117 \\
    ~~~~~~$\lambda=11.7$~\micron & 126 & 106 \\
    ~~~~~~$\lambda=12.4$~\micron & 138 & 107 \\
    $\chi^2_\mathrm{r}$ & 0.613 & 2.98 \\
    \hline
  \end{tabular}
  \tablefoot{$^\dagger$ Kept fixed during the fitting
    procedure.\\ Model~1 was fit to the visibilities only
    (Section~\ref{sec_tgm_vis}); Model~2 was fit to a combination of
    the visibilities and correlated fluxes
    (Section~\ref{sec_tgm_cflux}).}
\end{center}
\end{table}

\subsection{Model fits to the visibilities and correlated fluxes}
\label{sec_cflux_models}

The model fits to the visibilities presented in the previous section
are useful for gauging the appearance of IRAS~13481-6124 at different
wavelengths.  However, as mentioned in the previous section, fitting
the visibilities alone does not restrict the models (specifically, by
the temperature-gradient disk models) to be consistent with the SED.
As a next step, we embark on extracting some basic physical parameters
by means of comparing both the spatial \emph{and} spectral data to
more elaborate models.

Several previous studies of lower-mass, less-embedded stars utilize
similar models to successfully reproduce both the visibilities and the
entire SED \citep[e.g.][]{Malbet05,Eisner07,Ragland12,Vural12,Wang12}.
However, this technique is not directly applicable to deeply-embedded
objects, as the SED is dominated by the circumstellar envelope (rather
than the central star and disk) at all wavelengths.  At optical and
near-infrared wavelengths, and in the $N$-band within the silicate
feature, the emission from the star and/or disk is heavily absorbed
and reddened by the envelope.  At far-infrared/sub-millimeter
wavelengths, the emission from the envelope itself, which typically
contains the vast majority of the mass of the system, dominates the
SED.

In principle, the extinction caused by the envelope can be accounted
for at near- and mid-infrared wavelengths by reddening the model flux
values (or, equivalently, by dereddening the observed fluxes).
However, as mentioned in Section~\ref{sec_ext}, this cannot be done
reliably at near-infrared wavelengths (see also the detailed
discussion in Section~\ref{sec_dis_RT}).  On the other hand, the
optical depth derived in the $N$-band, which is dominated by the
silicate feature, is essentially independent of the type of
carbonaceous grains or the size distribution used, suggesting the
determination of the (total) extinction at these wavelengths to be
relatively robust.

We therefore use the following approach for the model fitting in this
section.  We fit the dereddened $N$-band correlated flux measurements
(the measured correlated fluxes were dereddened using the $R_V=5.5$
extinction law derived in Section~\ref{sec_ext}), and, at the same
time, the $K$-band visibilities (which are unaffected by extinction).
This hybrid approach of using correlated flux levels and visibilities
together allows us to make the best usage of the information available
in the observations, without over-interpreting the near-infrared flux
measurements by imposing a highly-uncertain near-infrared extinction
correction.  As in Section~\ref{sec_tgm_vis}, both the spatial
information and wavelength coverage are still represented in the
fitting process, but now the energy budget is also constrained by the
mid-infrared measurements.

\subsubsection{Temperature-gradient disk models}
\label{sec_tgm_cflux}

\begin{figure*}
  \begin{center}
    \includegraphics[width=170mm]{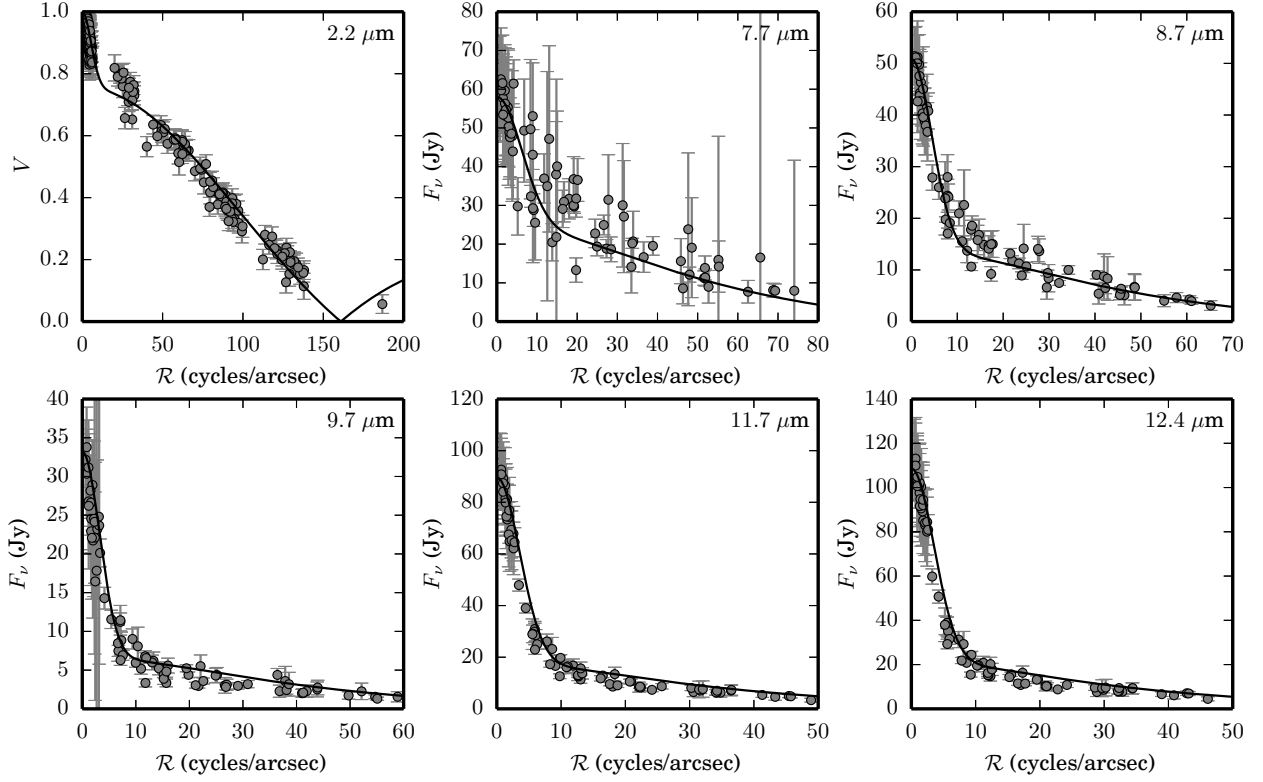}
    \caption{Temperature-gradient disk model fits to the $K$-band
      visibilities and $N$-band correlated fluxes (with
      $T_\mathrm{in}$ treated as a free parameter), plotted as a
      function of deprojected spatial frequency.  See
      Table~\ref{tab_tgm} (Model 2) for the fit parameters.  Circles and error
      bars mark the observations, while the model fits
      to each wavelength are shown as a solid curve.}
    \label{fig_tgm_cflux}
  \end{center}
\end{figure*}

\begin{figure}
  \begin{center}
    \includegraphics[width=85mm]{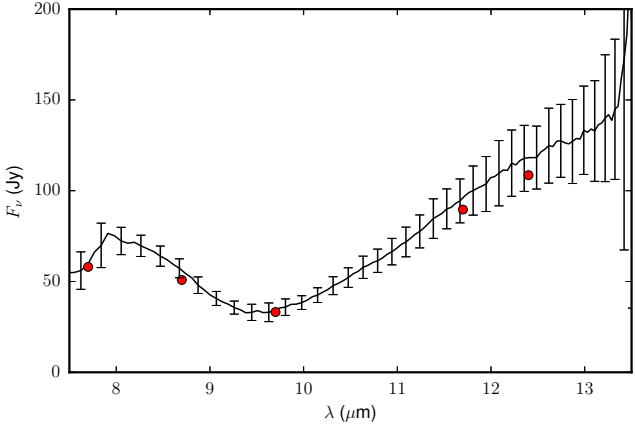}
    \caption{Total $N$-band flux levels for the model fit to the
      $K$-band visibilities and $N$-band correlated fluxes.  The flux
      levels predicted by the model are shown as red points.  The
      black curve and error bars show the MIDI spectrum, observed with
      the \textbf{prism}.}
    \label{fig_tgm_cflux_spectrum}
  \end{center}
\end{figure}

Applying this combined approach to the temperature-gradient disk model
presented in Section~\ref{sec_tgm_vis} allows us to leave the inner
temperature of the disk $T_\mathrm{in}$ as a free parameter.  We
present the results of the fitting procedure in Table~\ref{tab_tgm}
(Model~2), Fig.~\ref{fig_tgm_cflux} (interferometric data) and
Fig.~\ref{fig_tgm_cflux_spectrum} (total $N$-band spectrum).

Given the simplicity of this model and the wide range of wavelengths
and baselines fit, we regard this fit as good, despite having a larger
$\chi^2_\mathrm{r}$ value of 2.98.  In particular, we note that the
$N$-band correlated fluxes at long baselines, where the disk emission
dominates, are well reproduced by this model, despite the assumption
that the disk emits as a black body (i.e. it is not necessary to
include a silicate emission feature to reproduce the observed
correlated flux levels).  As with Model~1, the derived position angle
and inclination angle of the disk are perpendicular to the large-scale
outflow.  We derive a temperature of 1760~K for the inner disk, which
is fully consistent with the expectation that the inner radius is set
by the dust evaporation temperature.  As in the case of Model~1, the
majority of the \emph{total} mid-infrared flux originates from the
halo, suggesting a circumstellar disk makes only a relatively small
contribution to the total mid-infrared emission.

\subsubsection{Radiative transfer models}
\label{sec_rt_models}

\citet{Kraus10} presented a radiative-transfer model of
IRAS~13481-6124 based on available SED information and $K$-band
interferometric observations.  We note, however, that the modeling
efforts presented in that work do not reproduce the initial drop seen
in the visibility amplitude in the $N$-band at low spatial frequencies
(see Fig.~\ref{fig_kraus}).  As noted in Sec.~\ref{sec_visfits}, the
interferometric observations in the $N$-band show the presence of two
emission components, with the larger component (i.e., the ``halo'')
contributing most of the total flux.  This halo is also clearly
manifest in the $K$-band visibilities as a steep initial drop (e.g.
Fig.~\ref{fig_gaussfit}), which arises from scattered light from the
envelope.

In this section we explore the nature of these two separate emission
components, making use of radiative-transfer calculations.  For this
exercise, we use the radiative-transfer code developed by
\citet{Whitney03a,Whitney03b}, which is the same code used by
\citet{Kraus10} to model the $K$-band interferometric observations of
this object.  The radiative transfer code produces images and spectra,
which we convert into visibilities and correlated fluxes.  We use this
code for consistency with that work, and also because it allows the
inclusion of an envelope, with cavities carved out by a polar outflow.
Heated cavity walls have been claimed to be responsible for the
$N$-band emission at scales of $\sim100$~AU in other MYSOs
\citep[e.g.][]{deWit10,deWit11}, and this could be a natural
explanation for the two-component nature of the emission.

\begin{table*}
\begin{center}
  \caption{Parameters of radiative-transfer models}
  \label{tab_rt}
  \begin{tabular}{c c c c c c}
    \hline \hline
    Model & $\dot{M}_\mathrm{env}$ & $M_\mathrm{disk}$ &
    $r_\mathrm{in}$ & $\rho_\mathrm{cav}$ & $\rho_\mathrm{amb}$ \\
    & (\Msun~yr$^{-1}$) & (\Msun) & (AU) & (g~cm$^{-3}$) &
    (g~cm$^{-3}$) \\
    \hline
    Disk only & --- & 20 & 9.7 & --- & --- \\
    Envelope only & $4 \times 10^{-4}$ & --- & --- & $7.7 \times 10^{-20}$ &
    $9.1 \times 10^{-21}$ \\
    Disk and envelope & $2.65 \times 10^{-4}$ & $2.5 \times 10^{-3}$ &
    6.4 & $7.7 \times 10^{-20}$ & $9.1 \times 10^{-21}$ \\
    \hline
  \end{tabular}
\end{center}
\end{table*}

\begin{figure*}
  \begin{center}
    \includegraphics[width=170mm]{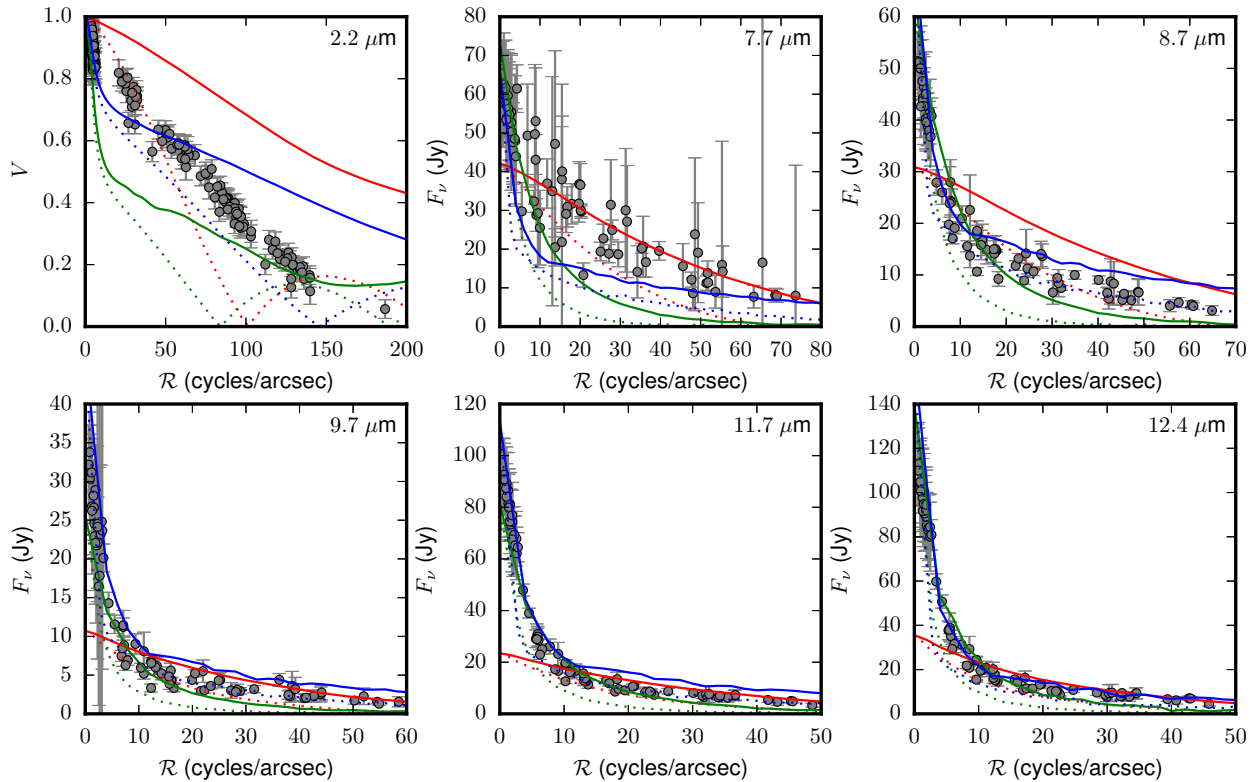}
    \caption{Visibility/correlated flux levels predicted by
      radiative-transfer models, plotted as a function of deprojected
      spatial frequency.  The observations are shown in gray.  The
      disk-only model is shown in red, the envelope-only model is
      shown in green, and the combined disk and envelope model is
      shown in blue, where the solid and dotted lines show the
      visibility perpendicular and parallel to the disk, respectively.
      See Table~\ref{tab_rt} for the fit parameters.}
    \label{fig_rt_models}
  \end{center}
\end{figure*}

As in Sec.~\ref{sec_tgm_cflux}, we limit ourselves to modeling the
$K$-band visibilities and the $N$-band correlated fluxes.  We do not
attempt to reproduce the entire SED with these models.  We explore
three classes of models, consisting of a disk only, an envelope (with
a 10\degr{} cavity) only, and a disk plus an envelope (also with a
10\degr{} cavity).  The disk structure is parameterized by a standard
accretion disk, with a scale height given by $H(r) = H_0
(r_\mathrm{in}/r)^\beta$ ($\beta=1.0$), and a vertically-truncated
inner rim.  The surface density distribution of the disk is adopted to
be a power law with an index of $-1.0$.  We fix the outer radius of
the disk to 1000~AU \citep[e.g.][]{deWit09}, and adjust both the inner
radius $r_\mathrm{i}$ and mass $M_\mathrm{disk}$ in our fits.  The
envelope structure is described by Eq.~(1) from \citet{Whitney03a},
and is parameterized by the mass-infall rate of the envelope
$\dot{M}_\mathrm{env}$, which we adjust as a fitting parameter, and
the centrifugal radius $R_\mathrm{c}$, which we also fix to 1000~AU.
The outer radius of the envelope was adopted to be 50\,000~AU.  The
values of $\dot{M}_\mathrm{env}$ quoted here merely parameterize the
density structure of the envelope, and do not represent a true
mass-infall rate.  We use the same dust for the disk and envelope
components (MRN-60Sil).  The dust opacity is set to zero in regions
where the dust temperature is found to be in excess of 1600~K.
Because it is not known in detail how the disk structure will change
in the presence of an infalling envelope, we adopt these
geometrically-simple structures.  For the central source, we adopt an
O8V star with $T_\mathrm{eff}=33400$~K and $R_*=8.5$~\Rsun{}
\citep{Martins05}.  The results of the three classes of
radiative-transfer models are shown in Fig.~\ref{fig_rt_models} and
summarized in Table~\ref{tab_rt}. 

In the disk-only model (red lines in Fig.~\ref{fig_rt_models}), we
have adopted the disk mass from \citet{Kraus10}.  In that work, they
calculated the formal dust sublimation radius to be between 6.2 and
10.9~AU, for an adopted luminosity of 34\,600~\Lsun{} (at a distance
of 3.2~kpc).  Scaling this to our adopted values by the square root of
the luminosity, we obtain a dust sublimation radius between 9 and
16~AU (or between 2.6 and 4.5~mas at a distance of 3.6~kpc).  For a
geometric ring model, the $K$-band visibilities indicate
$\theta=5.4$~mas \citep{Kraus10}, which translates to an inner disk
radius of 9.7~AU, and which we adopt as the inner radius for the
disk-only model\footnote{We note that this value is significantly
larger than the inner radius found for the temperature-gradient disk
models (see Table~\ref{tab_tgm}).}.  A circumstellar disk alone can
only account for the observed flux spectrum if additional dust
extinction ($A_V=50$~mag) is imposed, which provides an ad-hoc
approximation for the absorption due to the envelope.  The disk
emission can be made to fit the high-spatial-frequency tail in the
$N$-band correlated fluxes, however the correlated flux predicted at
low spatial frequencies is significantly lower than the observed
values.  Furthermore, the initial drop seen in the observed $K$-band
visibilities is also not present.  In other words, the disk-only model
cannot reproduce the larger-scale emission seen in the object.  We
experimented with the hydrostatic equilibrium solution in order to
increase the index $\beta$ (and, correspondingly, the disk flaring
angle) in order to make the disk brighter on larger ($\ga 50$~AU)
spatial scales, however, this did not improve the fit.  We underline
that, according to standard disk theory and hydrostatic equilibrium,
the index of the surface-density power law should lie between $-1.0$
and $-1.5$ \citep[e.g.][]{dAlessio98,Armitage10}.

We also explored the possibility of whether or not the interferometric
observations could be reproduced using a model consisting only of an
envelope with a cavity, without any contribution from a disk and
without additional foreground absorption applied.  This was the
approach used in previous studies of some MYSOs
\citep[e.g.][]{deWit10}.  For this envelope-only model (green lines in
Fig.~\ref{fig_rt_models}), we find that we can reproduce the emission
at low spatial frequencies, but the correlated flux levels at high
spatial frequencies are far too low.  The $K$-band visibilities are
also significantly lower than those observed.  In other words,
contrary to the disk-only model, the extended emission is reproduced
well, but there is far too little emission at scales of a few tens of
AU.  Clearly, one can anticipate that a combination of an envelope and
a disk will be able to reproduce the observed correlated flux levels
at both scales.

The combined disk and envelope model is shown in blue in
Fig.~\ref{fig_rt_models}.  Here, we have decreased the mass of the
disk from 20~\Msun{} to just $2.5 \times 10^{-3}$~\Msun{}, which has
negligible effect on the visibility and correlated flux levels, as the
disk is optically thick at infrared wavelengths in both cases.  The
inner radius of the disk was decreased to 6.4~AU, i.e. the inner
radius of our best fit to the $K$-band visibilities and the $N$-band
correlated fluxes using the temperature gradient disk (Model~2 in
Table~\ref{tab_tgm}; see Sec.~\ref{sec_tgm_cflux}).  Although the
final model produces an acceptable fit over all size scales and nearly
the full $N$ band, it is clear that at the short wavelength edge
($7.7$~\micron{}) the combined disk and envelope model produces fluxes
lower than those observed.  Nonetheless, comparing the model
visibilities to those obtained in the $K$ band, we find a reasonable
correspondence for the combined disk and envelope model.  Although the
spread in visibilities between position angles parallel and
perpendicular to the disk is larger than observed, the initial drop in
visibilities due to the envelope compares well, which implies that
some of the asymmetries seen in the $K$-band disk image (related to
the disk rim) are in reality much smaller than in the dust-only
radiative-transfer models.

\section{Discussion}
\label{sec_discussion}

\subsection{Summary of modeling results and the structure of
  circumstellar material around IRAS~13481-6124}

In this work, we have used a variety of progressively-complicated
approaches to model the interferometric observations of
IRAS~13481-6124, all of which attempt to unravel the distribution of
the circumstellar dust surrounding the object.

The Gaussian fits to the visibility (Sec.~\ref{sec_analysis_gauss}) at
each wavelength all indicate the presence of a compact ($5-15$~mas)
component, elongated approximately along the expected direction of the
disk (i.e., perpendicular to the large-scale outflow reported by
\citet{Kraus10}).  Besides the compact component, these models also
include a more extended ``halo'' component, which spans $75-120$~mas,
depending on wavelength.  The relative flux of this halo component
grows with wavelength, containing only 21\% of the total flux at
2.2~\micron{}, and a full 84\% of the total flux at 12.4~\micron{}
(see Table~\ref{tab_gaussfit}).

\citet{Kraus10} interpreted the compact, $K$-band emission as arising
from a circumstellar disk.  This position is supported by both
geometrical arguments (the elongated structure seen is oriented
perpendicular to a large-scale outflow) and the radiative-transfer
model presented in that work.  Both the orientation and the axial
ratio (given by 1/$\cos i$) of the compact component seen in the $N$
band are similar to those of $K$ band.  Thus, it is tempting to also
attribute this emission to a circumstellar disk.  However, both
numerical calculations by \citet{Zhang13} and $Q$-band
($\lambda=20$~\micron{}) observations of MYSOs by
\citet{Wheelwright12} indicate that outflow cavities probably play a
more significant role at mid-infrared wavelengths.  Therefore, caution
is warranted in interpreting our $N$-band interferometric
observations.

The geometric fits to the visibilities and correlated fluxes with the
temperature-gradient disk model, presented in Sec.~\ref{sec_tgm_vis}
and \ref{sec_tgm_cflux}, support the hypothesis that the
\emph{compact} emission seen at both wavelength ranges can be
explained by a single physical structure, in this case a
geometrically-thin, optically-thick circumstellar disk.  By fitting
the (dereddened) $N$-band correlated flux levels
(Sec.~\ref{sec_tgm_cflux}) with this simple model, we are able to
derive a value of 1760~K for the inner temperature of the disk,
consistent with expectations for a dusty disk.

The radiative transfer models employed in Sec.~\ref{sec_rt_models},
which we use to explore the effects of the outflow cavity on the
interferometric observations, also indicate that long-baseline
correlated fluxes arise specifically due to the disk (and not due to
the outflow cavity).  Furthermore, by allowing us to experiment with
different disk geometries (unlike the temperature-gradient disk), the
radiative transfer models show that the short-baseline measurements
\emph{can not} be reproduced by a disk alone, even by increasing the
flaring angle of the disk.

By including an envelope in the radiative transfer models, we are able
to reproduce the observed $N$-band correlated flux levels at short
baselines, as well as the initial drop seen in the $K$-band
visibilities.  These models also indicate that the vast majority of
the total flux at mid-infrared wavelengths arises due to the inner
edges of the envelope, rather than from the disk.  However, despite
this fact, a disk extending inwards to $\sim6$~AU is still required in
order to reproduce the long-baseline correlated flux levels at
mid-infrared wavelengths, as well as the relatively high $K$-band
visibilities.

We explored the effects of the cavity density ($\rho_\mathrm{c}$) and
the ambient density ($\rho_\mathrm{amb}$) (the minimum density
threshold) on the radiative-transfer model fits. For both cases with
an envelope, we adopted values of $\rho_\mathrm{c}=7.7 \times
10^{-20}$~g~cm$^{-3}$ and $\rho_\mathrm{amb}=9.1 \times
10^{-21}$~g~cm$^{-3}$. At low spatial frequencies, we found that the
cavity density itself does not matter, as long as it is substantially
less than that of the envelope and disk.  For high spatial
frequencies, increasing the cavity density slightly decreases the
correlated flux due to increased absorption at these scales.  The
ambient density was found to be degenerate with both the mass infall
rate and the envelope size, i.e. it is possible to construct many
models with the same total extinction by varying these parameters.

Among the proposed models for outflow creation, the wide-angle wind
\citep{Li96} and the jet bow-shock model \citep{Chernin95} explain the
observed outflow properties best \citep{Lee00,Lee01,Frank14}.
\citet{Lee01} showed that, for simulations of isothermal winds, a
post-shock number density of $10^3-10^4$ is seen.  For bow-shock
models, number densities of $10^4$ are produced in the post-shock
region.  The densities varies between the low- and high-velocity parts
of the outflow; however, for the scales the interferometer is
sensitive to ($\sim10-100$~AU), we assume that the cavity density
changes very little, and, based on previous work, we adopt a value of
$n=10^4$~cm$^{-3}$ ($2.3 \times 10^{20}$~g~cm$^{-3}$).

As pointed out by \citet{Kraus10}, the disk mass is not constrained by
the interferometric data.  Their estimate of the disk mass (20~\Msun)
was performed by varying the disk mass of the adopted envelope
description and observing the subsequent effects on the model SED.
Given that most of the disk mass is at cold temperatures, and that the
disk is optically thick at infrared wavelengths, (sub)mm observations
are required to make a more accurate estimate of the disk mass.
Indeed, we find that radiative-transfer models with a disk mass of
20~\Msun{} or $10^{-2}$~\Msun{} do not produce a significant
difference in the infrared correlated fluxes and visibilities.  The
calculated disk surface temperatures are similar in both cases (within
the first 100~AU), apart from the different hydrostatic equilibrium
solution, which produces a higher sublimation for the more massive
(denser) disk.  A disk more massive than 20~\Msun{} produces higher
temperatures farther out ($\ga100$~AU) at the disk surface, increasing
the total flux, but not the correlated flux MIDI is sensitive to.

Although \citet{Kraus10} used SED modeling to attempt to constrain the
disk mass, due to the inherent uncertainties in modeling the SEDs of
MYSOs (see Sec.~\ref{sec_ext}, and a more detailed discussion in
Sec.~\ref{sec_dis_RT}), we believe there is currently no way to
reliably (even approximately) determine the mass of the disk around
IRAS~13481-6124 using existing observational data.
High-angular-resolution ALMA observations will be able to estimate the
disk mass from the optically thin (sub)-millimeter continuum emission,
although it would still be necessary to discriminate between the disk
and envelope contributions.  The most promising approach, in our
opinion, would be to use spatially-resolved observations of gas lines
with ALMA in order to construct a kinematic model of the disk, and
determine the enclosed mass (including the stellar mass) as a function
of radius \citep[e.g.][]{Johnston15}.  Until this is done, for this
object and other similar MYSOs, the question of the disk masses around
such objects remains an open question.

\subsection{Prospects for future observations}

Infrared interferometry, be it in the near- or mid-infrared, is one of
only a few ways to observe systems like IRAS~13481-6124 with the
required spatial resolution ($\la10$~mas) to resolve the inner regions
of the dust disk.  While the three-telescope beam combiner of AMBER
was able to provide sufficient $uv$ coverage to allow \citet{Kraus10}
to reconstruct an image from the near-infrared interferometric
observations, achieving adequate $uv$ coverage with a two-telescope
instrument like MIDI (now decommissioned) is essentially impossible.
This is due to the large amount of observational time required to
observe a sufficient number of projected baselines, as well as the
inherent limits on the information which can be recovered from the
interferometric phase using a two-telescope instrument
\citep{Monnier07}.

The MATISSE instrument \citep{Lopez08}, which will be one of the
second-generation interferometric facilities at the VLTI, is expected
to come online in the near future.  This new instrument will not only
simultaneously utilize four telescopes (either the 1.8-m ATs or the
8.2-m UTs), but also open up the $L$ ($3-4$~\micron) and $M$
($4-6$\micron) bands for interferometric observations, in addition to
providing full, simultaneous coverage of the $N$ band.  As a result, a
single observation with MATISSE will provide spectrally-resolved
visibility amplitudes for six projected baselines, as well as four
determinations of the closure phase, simultaneously in two wavelength
bands (either $L$+$N$ or $M$+$N$).  Using MATISSE, it will therefore
be feasible to create a model-independent image reconstruction of the
source at a variety of mid-infrared wavelengths, much as
\citet{Kraus10} did in the $K$ band using AMBER, rather than being
limited to the simple geometric models presented in this paper.

\begin{figure}
  \begin{center}
    \includegraphics[width=85mm]{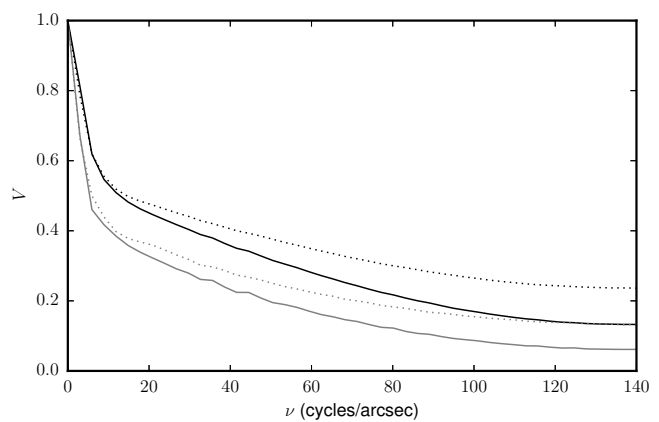}
    \caption{Predicted visibility levels in the $L$ (black) and $M$
    (gray) bands.  The solid and dotted lines show the visibility
    perpendicular and parallel to the disk, respectively.}
    \label{fig_matisse_V}
  \end{center}
\end{figure}

In anticipation of these new observational possibilities, we use the
combined disk and envelope radiative-transfer model from
Sec.~\ref{sec_rt_models} to predict the visibility levels in the $L$
and $M$ bands, which we show in Fig.~\ref{fig_matisse_V}.  We see that
the behavior of the visibility in both wavelength ranges can be split
into three spatial frequency ranges.  For low spatial frequencies
($\la10$~cycles/arcsec, $V\ga0.5$), we see that the envelope still
dominates, containing half of the total flux.  For spatial frequencies
in the range of $\sim20-120$~cycles/arcsec ($0.2\la V \la 0.5$), the
disk becomes evident, with a significant variation in the visibility
level for position angles perpendicular and parallel to the disk.
Finally, at high spatial frequencies ($\ga120$)~cycles/arcsec, the
visibility flattens off due to the unresolved central source.

In terms of projected baselines, the disk will be optimally resolved
for baselines of $7-90$~m in the $L$ band, and baselines of $10-120$~m
in the $M$ band.  This puts the disk just beyond the diffraction limit
of single $10$-m class telescopes, but perfectly in the range of
projected baselines available at the VLTI (currently up to
$\sim128$~m).  Given that the source is also very bright at these
wavelengths \citep[$3.0$~mag in the 3.4~\micron{} W1 filter and
$1.3$~mag in the 4.6~\micron{} W2 filter of WISE;][]{Cutri12}
IRAS~13481-6124 therefore represents an ideal target for MATISSE.

\subsection{On radiative-transfer modeling of deeply-embedded sources}
\label{sec_dis_RT}

\begin{figure}
  \begin{center}
    \includegraphics[width=85mm]{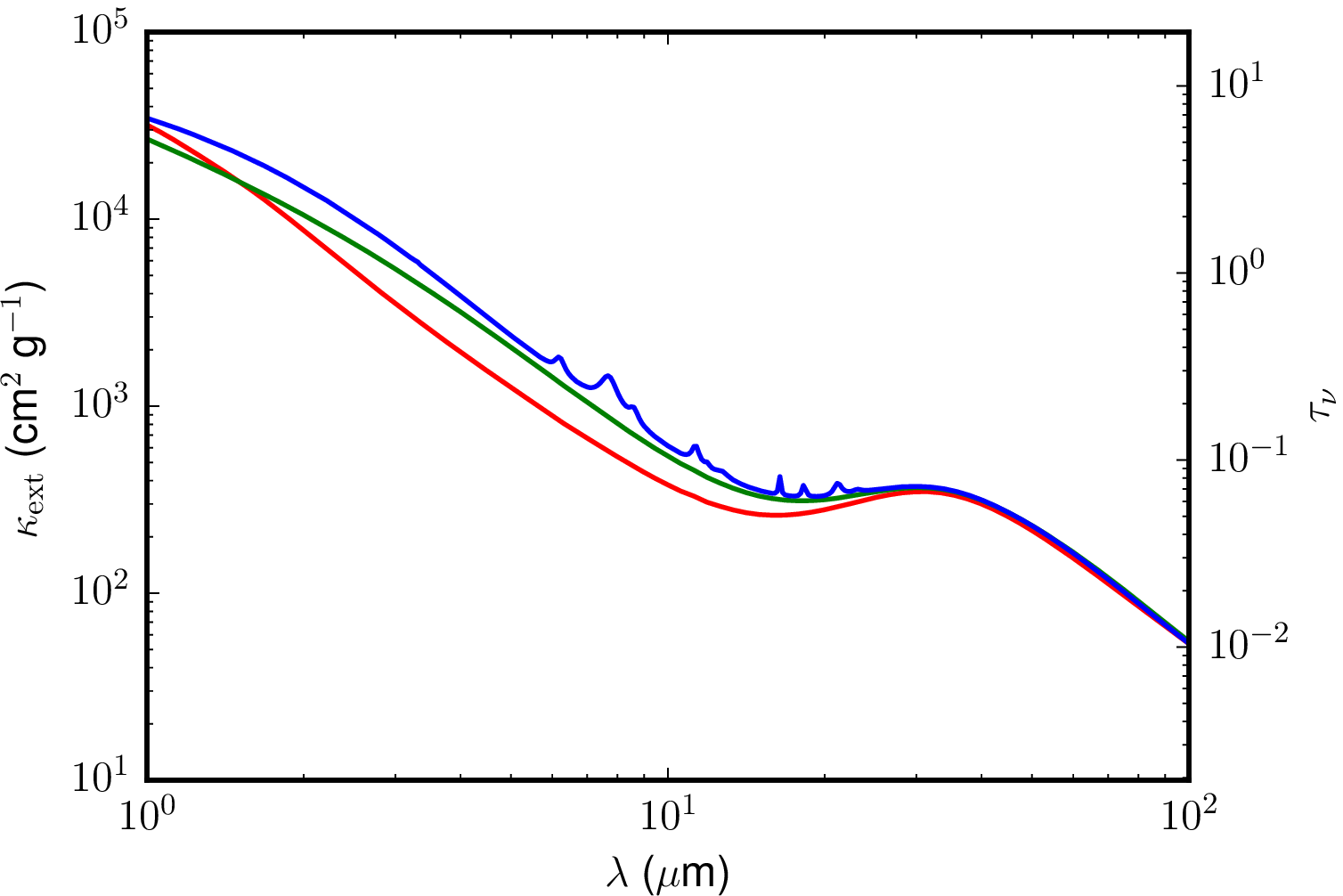}
    \caption{Near- and mid-infrared extinction derived for
      carbonaceous grains, using the size distributions of
      \citet{MRN77}, shown in red; \citet{Kim94}, shown in green; and
      \citet{WD01}, shown in blue.  The vertical axis shows the mass
      extinction coefficient on the left and the optical depth on the
      right, where a value of $3.8\times10^{-4}$~g~cm$^{-2}$ has been
      used for the column density of the carbonaceous material (see
      Section~\ref{sec_dis_RT}).}
    \label{fig_Cext}
  \end{center}
\end{figure}

At near-infrared wavelengths, where much of the emission from the
circumstellar disk is expected to originate, extinction from the
circumstellar envelope is already significant.  As discussed in
Section~\ref{sec_ext}, the amount of extinction is strongly dependent
on the type of carbonaceous grains and size distribution adopted.  Due
to the lack of spectral features at infrared wavelengths, which might
otherwise be used to constrain the composition and size distribution
of carbonaceous grains, it is impossible to reliably predict the
amount of extinction caused by the envelope, and therefore the amount
of emission from the circumstellar disk, at wavelengths
$\la8$~\micron{}.

We illustrate this point in Fig.~\ref{fig_Cext}, where we show the
infrared extinction properties obtained for carbonaceous dust for
three commonly-used grain size distributions: the classical MRN
power-law distribution \citep{MRN77}; the parameterized distribution
of \citet[hereafter ``KMH'']{Kim94}, which is a power law with
exponential decay for large grains; and the more elaborate ``Milky
Way'' size distribution of \citet[hereafter ``WD'']{WD01}.  In all
three cases, we have used the graphite grain properties from
\citet{Laor93}; for the WD size distribution, we also use the PAH
grains of \citet{Li01} for the small grains, following the
prescription given by \citet{WD01}\footnote{At these wavelengths, the
presence or absence of PAH grains does not affect the continuum
opacities, and the only effect of the PAH grains on the derived
extinction are the mid-infrared PAH bands.  Therefore, PAHs cannot be
used to restrict the infrared continuum extinction.}.  We show the
mass extinction coefficient $\kappa_\mathrm{ext}$, and also the
extinction optical depth $\tau_\nu$, where we have adopted a column
density of the carbonaceous material of
$3.8\times10^{-4}$~g~cm$^{-2}$.  This value was chosen based on the
estimate of $5.7\times10^{-4}$~g~cm$^{-2}$ for the silicate column
density by \citet{Boley13}, assuming a standard mass ratio of 60:40
for silicates and graphite \citep{Dwek97}.

The three extinction curves in Fig.~\ref{fig_Cext} represent three
reasonable, equally fair (given the information available) ``best
guesses'' for the contribution of carbonaceous grains to the dust
composition of the envelope.  However, the choice between these three
(or any other) size distributions/extinction curves will
\emph{strongly} influence the extinction at short wavelengths.  For
example, in the $K$ band ($\lambda=2.2$~\micron{}), the optical depths
$\tau_\nu(2.2$~\micron$)$ for the extinction curves constructing using
the MRN, KMH and WD size distributions are 2.7, 3.5 and 4.8,
respectively.  Consequently, the reddened $K$-band flux values (which
scale as $e^{-\tau_\nu(\lambda)}$) will be smaller by factors of 15
(MRN), 32 (KMH) and 130 (WD).  In other words, for deeply-embedded
sources, even the \emph{order of magnitude} of the amount of
attenuation from the envelope at near-infrared wavelengths is
difficult to determine.  Clearly, this is a very large uncertainty
when attempting to (simultaneously) model the emission from hot dust
at these wavelengths\footnote{To complicate matters further,
compositions, size distributions and optical properties for the
\emph{emitting} material must also be adopted, and these suffer from
many of the same uncertainties already noted here.}.

\section{Summary and conclusions}

In this work, we have presented new $N$-band interferometric
observations of the massive young stellar object IRAS~13481-6124,
obtained using aperture masking with T-ReCS on Gemini South, and
long-baseline measurements with MIDI on the VLTI.  We have examined
the extinction in the wavelength range of $\sim1-13$~\micron{} for
this source, and performed a comprehensive analysis of interferometric
observations of the object in both the $K$ and $N$ band, using both
geometric and radiative-transfer models.

On the basis of the geometric models, we have shown that both the
$K$-band visibilities and the $N$-band correlated fluxes can,
simultaneously, be well fit by a system consisting of a
geometrically-thin, optically-thick circumstellar disk, and an
extended halo.  In the $K$ band, the disk is the dominant source of
emission, producing $\sim76$\% of the total flux.  In the $N$ band,
however, the situation is reversed, with the disk contributing only
$\sim20-43$\% of the total flux, depending on wavelength.  For our
preferred model, the disk is inclined at $\sim48$\degr{}, and the
position angle of the semi-major axis is 107\degr{}.  We find the
inner radius of the disk to be at $\sim1.8$~mas, with a temperature of
1760~K.  The temperature gradient of the disk is well described by a
power law with an index of $-0.84$.

Using radiative transfer models, we have confirmed that a
circumstellar disk, by itself, is unable to produce the high
visibilities/large correlated-flux values observed in the $N$ band at
short baselines.  Furthermore, we found that the extended emission in
the $N$ band (i.e., the ``halo'' in the geometric models) can be well
reproduced by an outflow cavity, implying that the vast majority of
the $N$-band emission arises from the outflow cavity, rather than the
disk.  We have also shown that the infrared emission from the disk is
optically thick, and that the disk emission can be equally well
explained with disk masses ranging from $\sim10^{-3}$ to
$\sim10^{1}$~\Msun{}.

By comparing various extinction laws and considering their
applicability to this deeply-embedded object, we have shown that the
near-infrared absolute flux levels are completely unconstrained.  This
is due to the uncertain carbonaceous content of the envelope, which
dominates the extinction law at $\lambda \lesssim 8$~\micron.  The
near-infrared SED therefore contains very little (if any) information
about the disk.  In light of this, since previous determinations of
the disk mass have relied on modeling the SED, we argue that the disk
mass of IRAS~13481-6124, and, indeed, for other similar
deeply-embedded massive young stellar objects, cannot be determined
without spatially-resolved observations in the (sub-)millimeter
continuum and/or the disk kinematics.  On the other hand, we have
shown that both the near-infrared \emph{visibilities} and
(particularly at long baselines) mid-infrared \emph{correlated fluxes}
can be used together to very effectively probe the disk geometry.

\begin{acknowledgements} We thank the anonymous referee and editorial
staff of Astronomy and Astrophysics for thoughtful critique, which led
to the improvement of this manuscript.  The work of PAB was supported
by the Russian Science Foundation, grant No.~15-12-10017.  SK
acknowledges support from an STFC Ernest Rutherford fellowship
(ST/J004030/1) and a Marie Sklodowska-Curie CIG grant (SH-06192).  SL
acknowledges support from ANR-13-JS05-0005 and ERC-STG-639248.  JDM
acknowledges support from NSF-AST1210972.  We also thank Charlie
Telesco, David Ciardi, Chris Packham, Tom Hayward, Adwin Boogert, Jim
de~Buizer and Kevin Volk for their important contributions to the
Gemini/TRECS aperture masking experiment.  Some observations contained
herein were obtained at the Gemini Observatory (program~ID
GS-2007A-Q-38), which is operated by the Association of Universities
for Research in Astronomy, Inc., under a cooperative agreement with
the NSF on behalf of the Gemini partnership: the National Science
Foundation (United States), the Science and Technology Facilities
Council (United Kingdom), the National Research Council (Canada),
CONICYT (Chile), the Australian Research Council (Australia),
Minist\'erio da Ci\^encia e Tecnologia (Brazil) and Ministerio de
Ciencia, Tecnolog\'ia e Innovaci\'on Productiva (Argentina).
\end{acknowledgements}

\bibliography{refs}

\appendix

\section{Geometric fits to the $K$-band visibilities}
\label{sec_K_refit}

In this section, we provide an updated version of Table~S3 from
\citet{Kraus10}.  Due to an error in the program used in that work to
calculate the visibilities of the temperature-gradient disk model, the
outer annuli received too little weight.  As a result, the fitting
algorithm compensated by increasing $r_\mathrm{in}$.

\begin{table*}
\begin{center}
  \caption{Fit parameters for geometric models to $K$-band visibilities}
  \label{tab_K_refit}
  \begin{tabular}{l|c c c c c c|c c c}
    \hline \hline
    Model & $\theta$ & $r_\mathrm{in}$ & $T_\mathrm{in}$ & $q$ & $i^{(a)}$ & $\phi^{(b)}$ & $\chi^2_\mathrm{r,V}$ & $\chi^2_\mathrm{r,\Phi}$ & $\chi^2_\mathrm{r}$\\
    & [mas] & [mas] & [K] & & [\degr] & [\degr] \\
    \hline
    UD & 8.03 & & & & 41 & 111 & 4.90 & 2.59$^{(d)}$ & 4.49 \\
    RING & 5.38 & & & & 40 & 115 & 6.61 & 2.59$^{(d)}$ & 5.89 \\
    GAUSS & 5.43 & & & & 45 & 114 & 2.54 & 2.59$^{(d)}$ & 2.55 \\
    DISK~1500~K & & 1.27 & 1500$^{(c)}$ & 0.43 & 46 & 114 & 1.27 & 2.60$^{(d)}$ & 1.51 \\
    DISK~1800~K & & 1.14 & 1800$^{(c)}$ & 0.47 & 46 & 114 & 1.26 & 2.60$^{(d)}$ & 1.49 \\
    DISK~2000~K & & 1.12 & 2000$^{(c)}$ & 0.50 & 46 & 114 & 1.25 & 2.60$^{(d)}$ & 1.48 \\
    \hline
  \end{tabular}
  \tablefoot{As described in Appendix~\ref{sec_K_refit}, all models
    also include an extended component (Gaussian FWHM 108~mas) in
    order to reproduce the bispectrum speckle observations. \\($a$)
    This column gives the disk inclination, measured from the polar
    axis (i.e. 0\degr{} is pole-on).\\($b$) This column gives the model
    orientation, measured east of north.\\($c$) In the fitting
    procedure, this parameter was kept fixed.\\($d$) Given that these
    geometric models are point-symmetric, they predict a zero closure
    phase signal, resulting in the given $\chi^2_\mathrm{r,\Phi}$
    value.}
\end{center}
\end{table*}

We refer to Section~S4 of the work from \citet{Kraus10} for a detailed
description of the various models.  As in that work, we adopt an
extended component in the form of a Gaussian with a FWHM of 108~mas,
which contributes 15\% of the total $K$-band flux.  We show the
results of the fitting procedure in Table~\ref{tab_K_refit}.

We find the same results as \citet{Kraus10} for the UD, RING and GAUSS
models, which were not affected by the aforementioned error.  However,
for the temperature-gradient disk models, the inner radii
$r_\mathrm{in}$ are significantly smaller, and the temperature
gradients (described by the power-law index $q$) are steeper than
those reported by \citet{Kraus10}.

\section{Supplementary material}

\begin{figure*}
  \begin{center}
  \includegraphics[width=170mm]{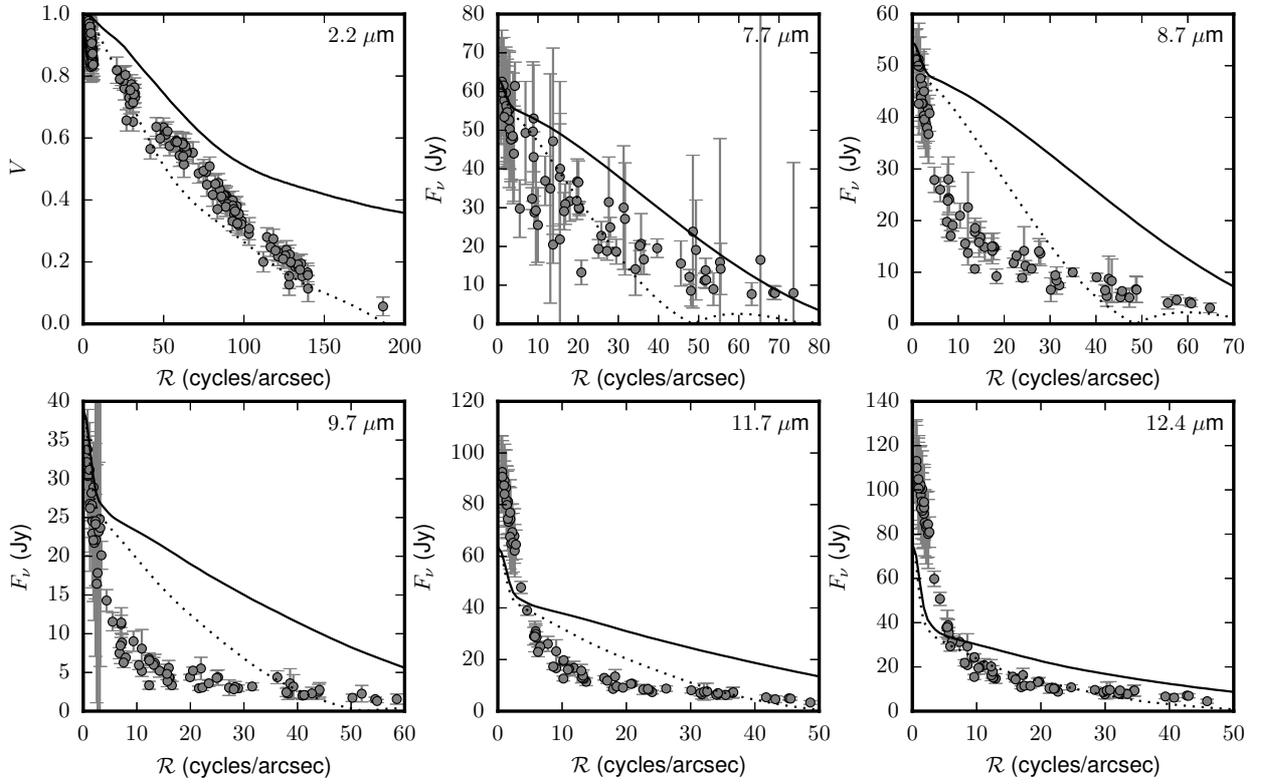}
  \caption{$K$-band visibilities and $N$-band correlated fluxes for
  the \citet{Kraus10} model, as a function of deprojected spatial
  frequency.  \textbf{The solid and dotted lines show the
  model predictions perpendicular and parallel to the disk,
  respectively.}}
  \label{fig_kraus}
  \end{center}
\end{figure*}

\end{document}